\documentclass[%
 reprint,
superscriptaddress,
 amsmath,amssymb,
 aps,
prapplied,
]{revtex4-2}

\usepackage{graphicx}
\usepackage{dcolumn}
\usepackage{bm}
\usepackage{physics}
\usepackage{amsmath,amssymb}
\usepackage{subfiles}
\usepackage{xcolor}
\usepackage[caption=false]{subfig}

\begin{document}


\title{Opportunities and Challenges of Computational Electromagnetics Methods for Superconducting Circuit Quantum Device Modeling: A Practical Review}

\author{Samuel T. Elkin}
\email{These authors contributed equally to this work.}
\affiliation{%
Elmore Family School of Electrical and Computer Engineering, Purdue University, West Lafayette, Indiana 47906, USA
}%
\author{Ghazi Khan}
\email{These authors contributed equally to this work.}
\affiliation{%
Elmore Family School of Electrical and Computer Engineering, Purdue University, West Lafayette, Indiana 47906, USA
}%
\author{Ebrahim Forati}
\affiliation{%
Google Quantum AI, Santa Barbara, California 93111, USA
}%
\author{Brandon W. Langley}
\affiliation{%
Google Quantum AI, Santa Barbara, California 93111, USA
}
\author{Dogan Timucin}
\affiliation{%
Google Quantum AI, Santa Barbara, California 93111, USA
}
\author{Reza Molavi}
\affiliation{%
Google Quantum AI, Santa Barbara, California 93111, USA
}
\author{Sara Sussman}
\affiliation{%
Fermi National Accelerator Laboratory, Batavia, Illinois 60510, USA
}%
\author{Thomas E. Roth}%
\email{rothte@purdue.edu}
\affiliation{%
Elmore Family School of Electrical and Computer Engineering, Purdue University, West Lafayette, Indiana 47906, USA
}%
\affiliation{Purdue Quantum Science and Engineering Institute, Purdue University, Indiana 47907, USA}




\date{\today}

\begin{abstract}
High-fidelity numerical methods that explicitly model the physical layout of a device play an essential role in the engineering and design of a wide range of technologies. For methods that are meant to characterize electromagnetic effects, these high-fidelity numerical methods are typically referred to as computational electromagnetics (CEM) methods. Although the CEM research field is relatively mature, with many commercial tools available that are used daily by researchers and engineers around the world, emerging applications can still stress the capabilities of the main techniques in wide use today. The design of superconducting circuit quantum devices falls squarely in this category of challenging applications due in part to the unconventional material properties (e.g., of thin-film superconductors), as well as due to important features of the devices covering nanometer to centimeter length scales. Such multiscale devices can significantly stress many of the fundamental properties of CEM tools which can lead to an increase in simulation times, a loss in accuracy, or in extreme cases cause the numerical method to break down completely so that no solution can even be reliably found. While these challenges of modeling multiscale devices are well known and are being actively investigated by CEM researchers, practical knowledge about these issues is limited in the broader research community of users of these CEM tools. This review is meant to serve as a practical introduction to the fundamental aspects of the major CEM techniques that a researcher may need to choose between in attempting to model a device, as well as provide insight into what steps they may be able to take to alleviate some of their modeling challenges. Our focus is on highlighting the main important features and concepts without attempting to rigorously derive all the relevant details, which can be found in many excellent textbooks and articles. After covering the fundamentals of CEM methods, we discuss more advanced topics related to the challenges of modeling multiscale devices with specific examples from superconducting circuit quantum devices to illustrate the concepts. We conclude with a discussion on future research directions that we anticipate will be valuable for improving the ability of researchers to successfully design increasingly more sophisticated superconducting circuit quantum devices. Although our focus and examples are taken from this application area, researchers from other fields will still be able to benefit from the details discussed here, especially if their simulations typically involve multiscale structures or unconventional material properties.
\end{abstract}

\maketitle
\section{Introduction}
Superconducting circuits are one of the leading hardware platforms for many quantum information technologies \cite{krantz2019quantum,gu2017microwave,blais2021circuit}, having found an especially prominent use in quantum computing \cite{arute2019quantum,wu2021strong,jurcevic2021demonstration,google2025quantum,mohseni2024build}, but also finding relevance in a variety of other areas such as quantum sensing \cite{chen2023detecting,kristen2020amplitude,dixit2021searching,braggio2025quantum,linehan2025estimating}. Although impressive progress continues to be made, there are significant engineering challenges that must be faced as devices scale up to unlock new applications \cite{mohseni2024build}. To accommodate vastly increased numbers of qubits, new packaging and integration strategies will be needed \cite{brecht2016multilayer,huang2021microwave,kosen2022building,conner2021superconducting,rosenberg20173d,wang2022hexagonal,google2025quantum,karamlou2024probing}, as well as significant miniaturization to many components \cite{bardin201929,patra202019,aumentado2020superconducting,macklin2015near,gu2017microwave,ranzani2019circulators,muller2018passive,navarathna2023passive}. These changes will complicate the already prevalent issues of classical and quantum crosstalk and susceptibility to noise \cite{krantz2019quantum,ash2020experimental,tripathi2022suppression,chen2021exponential,jurcevic2021demonstration,acharya2023suppressing,miao2023overcoming,Fan2025calibrating}, which contribute to decoherence that must be suppressed. Further, the speed and fidelity of qubit control and readout must also be improved \cite{chen2021exponential,acharya2023suppressing,google2025quantum,bengtsson2024model}. While many factors influence these myriad considerations, electromagnetic (EM) effects that can be controlled and optimized by device design influence all of them. As a result, increasing interest is being placed in how we model these effects with corresponding calls for improvement in these modeling procedures to address the associated engineering challenges \cite{mohseni2024build,Levenson-Falk_2025,shanto2024squadds}. 

Fundamental to all of these modeling procedures is the use of some kind of computational electromagnetics (CEM) tool that takes as input a 3D representation of the physical device with the associated material properties and sources and then computes the corresponding EM effects according to some governing physical equation. These CEM tools have found use in many applications such as antenna and scattering design, signal integrity analysis in on-chip interconnects, geophysical modeling, semiconductor and photonic device design, among many others \cite{taflove2005computational,jin2011theory,jin2015finite,chew2001fast}. Correspondingly, CEM tools have reached a level a maturity where there are many commercial and open-source products available that implement different computational methods to choose from. Faced with this wide range of options, it can be difficult for newer users unfamiliar with the underlying principles of the different CEM methods to determine which modeling tool may be most appropriate for a particular task. These challenges are exacerbated in the area of modeling circuit quantum electrodynamics (cQED) devices \cite{blais2021circuit}, which have unique features that can often stress the limitations of conventional CEM implementations.

For instance, compared to traditional EM applications, cQED devices use unconventional materials such as thin superconductors that CEM tools have not been built to model. Further, the cryogenic temperatures lead to dielectric loss seemingly being dominated by thin interface layers rather than in the bulk of materials \cite{read2023precision}, causing users to need to consider methodologies to model the effects of nanometer-thick layers in complicated device layouts that can cover centimeters of space in the transverse dimensions. Even without considering these thin layers explicitly, typical transmission lines and qubits still have micrometer-scale features, leading to geometries with important dimensions spanning many orders of magnitude. Such multiscale devices are difficult to model with conventional CEM tools because of limitations in the discretization procedures, potentially causing slower simulation speeds and hard-to-diagnose inaccuracies in the models to occur.    

Given this wide range of modeling needs and associated challenges, over the years a multitude of methods focused on predicting aspects of cQED device performance using CEM tools have emerged. Early systematic efforts employed black-box quantization approaches \cite{nigg2012black,solgun2014blackbox}, which from a CEM perspective involve performing conventional driven simulations to compute the multi-port impedance matrix of a device (albeit, over a possibly unconventionally wide frequency range). More recently, efforts to explicitly utilize eigenmode computations from CEM tools have also gained attention \cite{minev2021energy,roth2021macroscopic,moon2024analytical}. Other simulation approaches have focused on ways to use impedance parameters (or other microwave network parameters) to directly compute quantum properties of interest, such as spontaneous emission rates or qubit-qubit exchange coupling rates \cite{houck2008controlling,solgun2019simple,roth2022full,solgun2022direct,khan2024field,labarca2024toolbox}. While all these examples use a full-wave CEM tool (i.e., one that incorporates wave physics through the complete set of Maxwell's equations), there are also common simulation uses of quasi-static CEM solvers to extract lumped element circuit representations of parts of devices \cite{Levenson-Falk_2025}. There can also be various use cases for simulations performed with a time domain solver, such as for performing multiphysics modeling of qubit control and readout \cite{roth2024maxwell,2024_Elkin_JTWPAs,elkin2025ims,elkin2025multiphysics} or as an alternative method to compute quantities like impedance parameters or eigenmodes over a broad band in a single simulation. Although we have focused our discussion here on cQED modeling, similar use cases can be found in the modeling of many other quantum information technologies as well \cite{moon2024computational}.

Despite the prominent role of CEM tools in cQED device design, most users treat them as a ``black-box'' with limited understanding of how the tools actually work. This can easily lead to misunderstandings that hamper the ability of researchers to use them effectively in their designs, particularly given that typical cQED devices stress some of the limitations of many of these tools. The purpose of this review is to help users of CEM tools achieve a more informed understanding of how the tools operate and what steps they can take to try and alleviate some of the modeling challenges. While our discussion and simulation examples will be focused on structures commonly encountered in cQED devices, the concepts will still be useful to many other application areas that also encounter multiscale features and unconventional material properties. We also identify a number of opportunities for further research on CEM tools and how to best use them for cQED modeling, which we hope can aid in closing the gaps between initial simulations and the corresponding measurements on a fabricated device. Closing these gaps has been critical to the success of a plethora of technologies, including the design of classical computer processors, and we expect this will be a key piece in the overall development of quantum information technologies as well.

The remainder of this work is organized in the following way. In Section \ref{sec:numerical_linear_algebra}, we review some basic concepts from numerical linear algebra that will support our later discussions on CEM methods. Following this, we present in Section \ref{sec:cem_overview} a broad overview of the main CEM methods one is most likely to encounter in their work with an emphasis on the basic details that allow us to convert from a continuous physical equation to a discretized matrix equation that can be solved on a computer. These discussions are then specialized in Section \ref{sec:eig-overview} to the additional challenges faced in computing a large number of eigenvalues and eigenmodes. We then discuss in Section \ref{sec:practical-considerations} more advanced topics that play an important role in understanding why cQED devices are hard to model and what a user of these tools may be able to do to improve the likelihood of a successful simulation.  Finally, we conclude in Section \ref{sec:conclusion} by discussing opportunities for future research to improve the modeling of cQED devices.

\section{Numerical Linear Algebra Primer}
\label{sec:numerical_linear_algebra}
To support our later discussions on the properties of typical CEM methods, it will first be helpful to review some central concepts from numerical linear algebra. As we will see in Section \ref{sec:cem_overview}, most CEM methods of interest for modeling cQED devices eventually require the solution of a matrix equation. Hence, here we will discuss the distinction between direct and iterative solvers for solving matrix equations and the concept of a condition number of a matrix. Further, we will discuss some of the basic terminology and concepts surrounding the solution of eigenvalue problems that arise in CEM. Readers already familiar with these concepts should be able to skip directly to Section \ref{sec:cem_overview}.

\subsection{Direct vs. Iterative Solvers}
\label{subsec:direct-vs-iterative}
A standard matrix equation given by
\begin{align}
    [A]\{x\} = \{b\}
\end{align}
can be solved using a wide array of possible numerical strategies. We generally classify these approaches into two different families: direct and iterative solvers.

\begin{figure}[t!]
    \centering
    \includegraphics[width=0.9\linewidth]{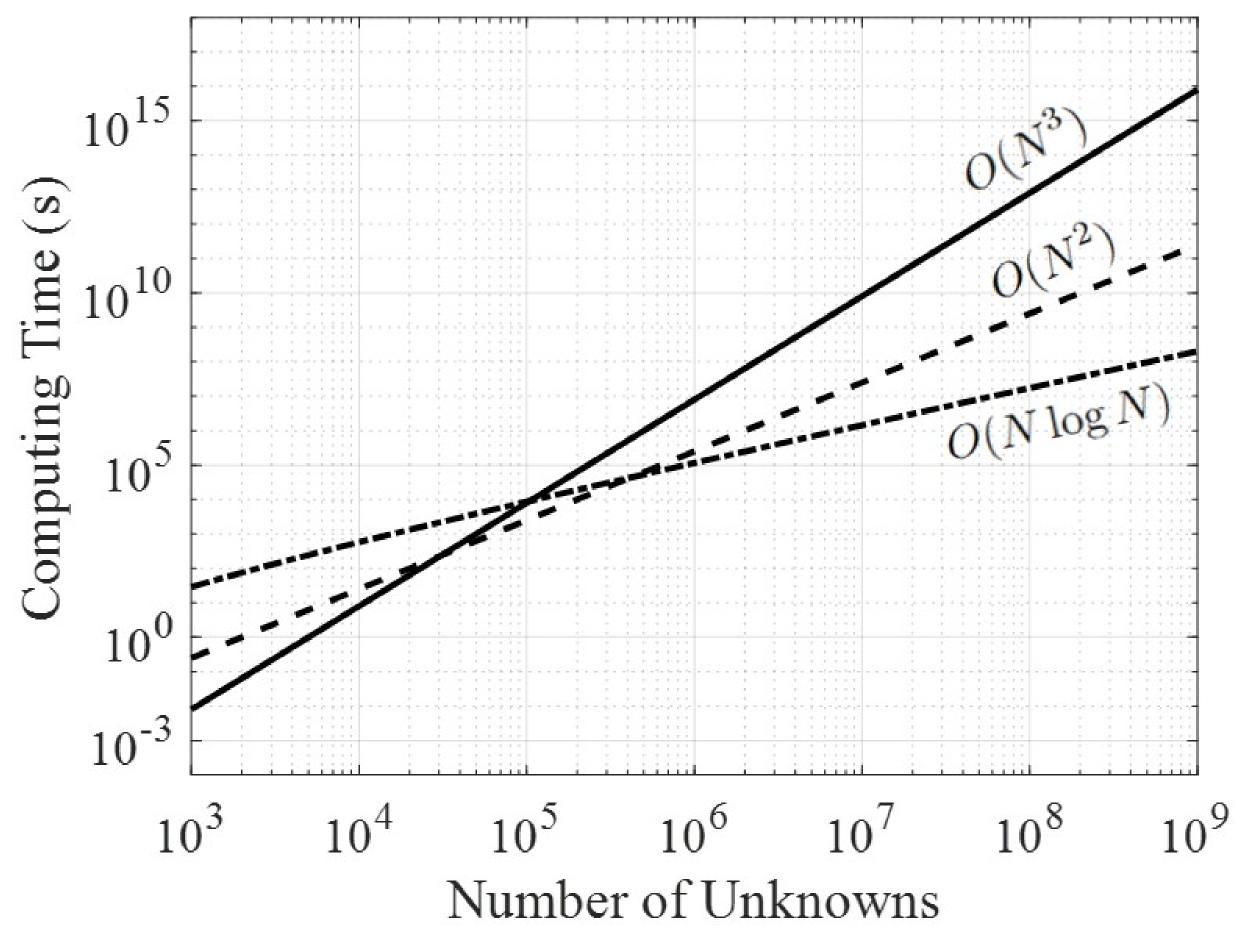}
    \caption{Notional computation times for different algorithms with varying computational complexities using a single-core processor (adapted from \cite{jin2011theory}). While parallelization can make solving problems with billions of unknowns possible, this only occurs for algorithms with suitable computational complexity, which is usually $\sim \! O(N\log N)$ for practical algorithms.}
    \label{fig:computational-scaline}
\end{figure}

A \textit{direct solver} either explicitly (or in essence) ``directly'' computes the inverse of the matrix. Examples of direct solvers are Gaussian elimination routines or performing a lower-upper (LU) decomposition; although, other more specialized and advanced approaches also exist for special matrix types like the sparse matrices (i.e., most of the matrix entries are zeros) that typically arise from discretizing partial differential equations (PDEs). The straightforward direct solver techniques such as Gaussian elimination or LU decompositions that make no assumptions about the matrix structure are typically only applicable for solving relatively small problems. The reason for this is the order of operations required to complete the numerical routine for these simple approaches is typically $O(N^3)$, where $N$ is the dimension of the matrix. This scaling can quickly lead to inordinate computation times that are completely unacceptable (e.g., see Fig. \ref{fig:computational-scaline}). 

Hence, there is significant research interest devoted to improving the speed of direct solvers for specialized problem sets. However, even these specialized techniques will typically still struggle as the system size grows large due to either the computation time or the memory requirements needed in the process. As a rough rule of thumb, general-purpose direct solvers tend to become difficult to use in the range of $10^4<N<10^5$. High-performance and parallelized implementations (e.g., \cite{bueler2020petsc}) for specialized matrices such as sparse ones can push beyond this (typically, by an order of magnitude or more depending on the computing hardware used), but ultimately direct solvers are still limited in the size of matrix system they can address.   

An alternative to a direct solver is to use an \textit{iterative solver}. These methods can be viewed as being similar to a kind of optimization routine. In their most basic form, they begin by guessing a trial solution $\{{x}_0\}$ and then compute $[A]\{x_0\}$. The residual error is computed as $\{r\} = [A]\{x_0\} - \{b\} $. This residual error is then used to update the trial solution to some new guess $\{x_1\}$. The exact way this residual error is utilized depends on the particular iterative solver employed, of which there are many options available; however, for EM applications, Krylov subspace based iterative solvers are the most practical. This process of using the residual error to improve our trial solution continues until we ideally reach some level of convergence (i.e., the residual error drops below a specified value). One of the main benefits of this process is that a core computational bottleneck of these methods becomes computing the matrix-vector product $[A]\{x_n\}$. In a worst-case scenario (i.e. a completely dense matrix), this operation can be completed in $O(N^2)$ steps, which can lead to a huge time saving compared to a direct solver (see Fig. \ref{fig:computational-scaline}), but is still not adequate for large problem sizes. When working with sufficiently sparse matrices (like those made from discretizing a PDE), the cost of a matrix-vector product can typically be completed in $O(N)$ operations, which becomes much more practical for solving large-scale simulation problems.  

However, some unique aspects of EM modeling problems still lead to some important challenges when using iterative solvers that must be addressed to successfully model large-scale problems. Of primary importance is that many EM problems can have slow convergence (which requires many iterations of the iterative solver), or in extreme cases may not be able to converge at all. This can complicate the use of the most popular and robust iterative solvers generally used in EM applications, the Krylov subspace based approaches. In some of the more robust and sophisticated Krylov subspace based algorithms, multiple vectors will be built up and processed collectively to aid in choosing better ``search directions'' to update the new solution guess $\{x_n\}$. For large-scale problems, the storage and processing of these vectors can become prohibitive, requiring the algorithm to be occasionally ``restarted'' by discarding the stored vectors and beginning the formation of a new set to continue the algorithm with. As a result, there is always interest in creating discretization approaches that will be more amenable to iterative solvers so that they can function in a more ideal and efficient manner. 

In EM applications, typical iterative solvers are the conjugate gradient (CG) method and biconjugate (BiCG) gradient method \cite{jin2015finite}. In some situations, the matrices involved will no longer satisfy the requirements to apply these CG-based methods. In these cases, it is common to turn to the generalized minimal residual (GMRES) method, which is a very robust iterative solver that does not have strict restrictions on the kinds of matrices it can be applied to \cite{saad1986gmres}. However, many other iterative solver options exist and can be exploited in certain scenarios.

\subsection{Matrix Condition Number and Preconditioning}
\label{subsec:condition-numbers}
When discussing iterative solvers in Section \ref{subsec:direct-vs-iterative}, we somewhat vaguely said that EM modeling problems can often have slow convergence. In numerical linear algebra, we typically quantify the ``difficulty'' in solving a problem in terms of the condition number of the matrix. Mathematically speaking, the condition number of the matrix is typically given by the ratio of the largest to smallest singular values of the matrix (although other definitions are sometimes also used). More intuitively, the condition number is a measure that can be conceptually thought of as being related to the sensitivity of the matrix equation to changes in the inputs. For instance, if a problem has a large condition number, it will be very sensitive so that even small changes in $\{b\}$ can lead to very large changes in $\{x\}$. This can correspondingly lead to difficulty in achieving convergence with an iterative solver, since effectively ``searching'' the solution space is complicated by the large sensitivity to changes.

More broadly, it is also common to see the condition number be related to the ``achievable accuracy'' in solving a problem. In these cases, it is generally given as a rule of thumb that for a condition number of $10^k$ one can expect to lose up to $k$ digits of accuracy in solving the problem due to errors associated with floating-point arithmetic. It should be noted that this loss of accuracy is only in the solution procedure itself, and is not accounting for the approximation errors that invariably are incurred when one discretizes a continuous problem to be solved on a computer (i.e., the errors due to meshing and other aspects of the physics are not being accounted for in this $k$ digit ``loss of accuracy'').

More colloquially, a problem is usually referred to as well-conditioned if one can expect an iterative solver to perform well and ill-conditioned if it will be more challenging to solve. Unfortunately, there are many aspects of practical EM problems that can lead to them being fairly ill-conditioned problems. As a result, it is imperative in most practical applications to utilize some kind of preconditioner to improve one's ability to successfully solve the matrix equation. A preconditioner is generally formulated to be some kind of simple approximation to the inverse of $[A]$ that can be incorporated into the matrix equation prior to attempting to solve it (in the simplest case, this could be left-multiplying by the preconditioner). If the preconditioner does a decent job approximating $[A]^{-1}$, the remaining ``job'' of the iterative solver is greatly simplified and quicker convergence to the solution can be achieved. In practice, there is always a problem-specific trade-off between how sophisticated of a preconditioner should be used; generally, a more complicated preconditioner will take longer to numerically form, but then lead to the iterative solver converging quicker. Generally, a more advanced preconditioner will be costlier than $O(N)$, but practical algorithms with $\sim\!O(N\log N)$ scaling do exist for this. Overall, there is always research going on for various applications on formulating specialized preconditioners for a particular application to achieve the best possible performance. 

In EM applications, various features will typically lead to a more ill-conditioned matrix. First, the basic structure of wave physics tends to lead to more ill-conditioned matrices than other physical problems (e.g., diffusion is generally simpler to deal with than wave propagation). Beyond this basic structural challenge, the conditioning of EM matrices that solve the wave equation tend to get worse as the frequency being solved at is lowered and/or as the mesh density is increased. These are often referred to as low-frequency and dense-mesh breakdowns, respectively. This is a consequence of aspects of the physics becoming more ``quasi-static'', which stresses certain terms in the wave equation to lead to disparate numerical scaling of constituent parts of the equations that makes the conditioning worse. These challenges generally get combined and exacerbated when modeling multiscale devices which contain both very small geometric features and larger parts (e.g., having extents over multiple wavelengths), as will be discussed more in Section \ref{subsec:multiscale}. As a result, some amount of preconditioning is almost always needed in solving EM problems, and more specialized approaches can be needed for particularly challenging application spaces and complex geometries.

\subsection{Eigenvalue Problem Basics}
\label{subsec:eigenvalue-problem-basics}
In addition to solving the ``standard'' matrix equation $[A]\{x\}=\{b\}$, another common class of problems in numerical linear algebra is the solution of eigenvalue problems. The standard eigenvalue problem is to solve
\begin{align}
    [A]\{v_n\} = \lambda_n \{v_n\},
    \label{eq:eigenvalue-problem}
\end{align}
where $\lambda_n$ is the eigenvalue and $\{v_n\}$ is the eigenvector, which collectively are referred to as an eigenpair. In finding eigenpairs related to physical PDEs, generally the discretization strategies employed lead to a \textit{generalized eigenvalue problem} of the form
\begin{align}
    [A]\{v_n\} = \lambda_n [B] \{v_n\}
    \label{eq:gen-eigenvalue-problem}
\end{align}
that must be solved instead of (\ref{eq:eigenvalue-problem}). For EM applications, the eigenvalue will be related to the resonant frequency of the modes of the structure being modeled and the eigenvector will correspond to the associated resonant field distribution. Unfortunately, although eigenvalue problems are of significant interest, they are also substantially more difficult to solve numerically than standard matrix equations.

As with solving standard matrix equations, it is possible to employ ``direct'' or iterative solvers to the solution of eigenvalue (or generalized eigenvalue) problems. There are some nuances to this, but for simplicity we will refer to a ``direct'' eigensolver as one that is formulated in such a way that it must compute all of the eigenvalues (or eigenpairs) of the matrix, while an iterative eigensolver will be able to compute some fixed subset of the total eigenpairs. Typically, these iterative solvers can be targeted to find eigenpairs with certain properties (e.g., the eigenpairs with the largest or smallest $m$ eigenvalues, or with eigenvalues closest to some specified ``target eigenvalue''). However, it should be noted that not all algorithms perform equally well at finding such specified eigenpairs. For example, it is easier to find the eigenpairs associated with the largest eigenvalues rather than the smallest. In most situations, using direct solvers for eigenvalue problems is extremely impractical unless the system being modeled is quite small (e.g., the matrix dimension is in the low $10^4$ range), although high-performance packages can help with tackling larger systems.

Natively, most eigenvalue algorithms are best suited for finding the largest eigenvalues of a matrix (and their associated eigenvectors). However, these eigenpairs are typically not of physical interest in EM applications, as we usually wish to locate the smallest eigenpairs or those beginning near a target frequency value. The standard solution to this problem is to operate the algorithm in ``shift-invert mode'', where we instead seek eigenpairs of the modified system $([A] - \sigma [I])^{-1}$, where $[I]$ is the identity matrix. By subtracting $\sigma [I]$, the entire spectrum of $[A]$ is shifted downwards by $\sigma$. As a result, if $\sigma$ is selected close to the eigenvalues of interest, these eigenvalues are shifted close to $0$. Inverting this system causes these very small eigenvalues to instead become very large, allowing the associated eigenpairs to be evaluated readily using standard methods. This procedure can be extended to generalized eigenvalue problems in the form of (\ref{eq:gen-eigenvalue-problem}) by searching for eigenpairs of $([A] - \sigma [B])^{-1}$.

In practice, software packages typically allow a user to specify a method of solving $([A] - \sigma [I]) \{x\} = \{b\}$ (or $([A] - \sigma [B]) \{x\} = \{b\}$ for a generalized eigenvalue problem) for an arbitrary vector $\{ b \}$. As discussed in Section \ref{subsec:direct-vs-iterative}, this can be accomplished using direct or iterative solvers. If possible, a direct factorization of the modified system is preferable, as many iterations of the eigenvalue solver are typically required, corresponding to solving the system for many vectors $\{b\}$. However, eventually computing these factorizations will require excessive amounts of time and memory, meaning an iterative solver must be used instead. In this case, the costs of repeated calls to this inner solver usually need to be mitigated by forming an effective preconditioner as a pre-processing step to minimize the number of iterations that are required for the inner solver to converge. Further, the convergence of the outer iteration for finding the eigenvalues is dictated by the spacing between the eigenvalues being sought. As a result, the number of outer iterations can be minimized if $\sigma$ is selected very close to the desired eigenvalues. Unfortunately, if the problem being solved has many ``clustered'' eigenvalues (i.e., eigenvalues that have similar values to one another), then the problem can remain inherently difficult to solve accurately.

From this discussion, one can surmise that there is significantly ``more'' involved in solving an eigenvalue problem than in solving a standard linear system of equations. This is unfortunately true, and solving large eigenvalue problems remains one of the most challenging tasks in numerical linear algebra \cite{jin2015finite,golub2000eigenvalue}. In particular, one of the central challenges is in calculating a ``large'' number of eigenpairs from a ``large'' matrix. What constitutes as ``large'' is application specific, but generally attempting to compute $O(10^3)$ eigenpairs from matrices of size $O(10^6)$ to $O(10^9)$ would often be incredibly challenging.

\subsection{Parallel Computing}
\label{subsec:parallel-computing}
The advent of parallel computing has significantly pushed what is possible in numerical linear algebra and is a core component of most modern algorithms. However, it is important to note that not all algorithms or CEM techniques are easily parallelizable, so significant effort can be required for parallel computing to address computing time bottlenecks. Different kinds of parallel architectures (e.g., CPU versus GPU or shared memory versus distributed memory) also require different algorithmic solutions. Due to the complexity and intricacy of these issues, we will not delve into them in depth here. However, we will note before moving on that parallelization is not a universal ``fix''; if an algorithm has a poor computational scaling like the $O(N^3)$ methods discussed previously, one will quickly find that no amount of practically-accessible parallelization will be able to ``save them'' for problems of large size that are usually of practical interest. Beyond computing time, a significant challenge and bottleneck for many parallel algorithms is the amount of computer memory available and quickly accessible within the computing architecture. This becomes a particular concern for strategies such as domain decomposition methods that will be discussed in Section \ref{subsec:domain-decomposition}.
\section{Overview of CEM Methods}
\label{sec:cem_overview}
In this section, we will discuss the main classes of CEM methods with a focus on ``driven simulations'' where there is an explicit source included in the simulation (e.g., a current or voltage source, an incoming EM wave, etc.). Further, our discussions in this section will primarily be on how the various methods convert a continuous problem into a discrete one that can be solved on a computer using numerical linear algebra techniques. Further applied concepts relevant to using these tools in the practical analysis of cQED devices will be covered in Section \ref{sec:practical-considerations}. Additionally, considerations for these different methods in the context of eigenvalue problems will be postponed to Section \ref{sec:eig-overview}. 

Here, we begin by discussing finite difference methods (FDMs) in Section \ref{subsec:fdm}, which are somewhat special in that they are mainly used for time domain analyses. We then discuss the finite element method (FEM) for solving PDEs in Section \ref{subsec:fem}. This is followed in Section \ref{subsec:mom} where the method of moments (MoM) is presented for solving EM integral equations. For each of these foundational techniques, we discuss their fundamental concepts and unique considerations for using them in time domain analyses.  Other kinds of asymptotic or hybrid methods that can be encountered are discussed in Section \ref{subsec:cem-other}. We also briefly discuss the main ideas of domain decomposition methods in Section \ref{subsec:domain-decomposition}. Finally, in Section \ref{subsec:cem-summary}, we summarize the main points about the different CEM methods and their relevance to typical cQED modeling applications.

\subsection{Finite Difference Methods}
\label{subsec:fdm}
The FDM approach to discretizing a PDE is the simplest option available, and as such is a very accessible computational technique to learn and implement. This has made it one of the most popular methods for use in research and in many other EM applications. However, this simplicity does come at a cost in terms of the geometrical fidelity and accuracy that can be achieved, which is discussed in more detail throughout this section.

\subsubsection{Fundamental Concepts}
\label{subsubsec:fdm-fundamentals}
Fundamentally, FDMs are formulated by approximating the derivatives that appear in a PDE (e.g., in Maxwell's equations) using an approximate form of the mathematical definition of a derivative. Typically, a \textit{central difference} approximation is used by expressing a derivative as
\begin{align}
    f^\prime\left(x\right)\approx\frac{f\left(x+\Delta x\right)-f\left(x-\Delta x\right)}{2\Delta x},
	\label{eq:derivative-definition}
\end{align}
where $\Delta x$ is some ``small'' step size relative to the expected variations in $f$. As $\Delta x$ becomes smaller, the approximation becomes more accurate. This same central difference approximation can be applied again to the above equation for $f^\prime\left(x\right)$ to arrive at an approximation for higher-order derivatives.

To see how FDMs can lead to a matrix equation, we will briefly consider the simple case of Poisson's equation in 2D with a homogeneous permittivity. In this case, Poisson's equation can be written for the scalar potential $\phi$ as
\begin{align}
    \partial_x^2\phi + \partial_y^2 \phi = -\rho/\epsilon,
\end{align}
where $\rho$ is a prescribed charge density. If we use identical grid sizes in $x$ and $y$ as $\Delta x = \Delta y = h$, we can apply central difference approximations at the grid point $(i,j)$ to get
\begin{multline}
    \frac{\phi(i+1,j)-2\phi(i,j) + \phi(i-1,j)}{h^2} \\ + \frac{\phi(i,j+1)-2\phi(i,j)+\phi(i,j-1)}{h^2} = -\frac{\rho(i,j)}{\epsilon},
\end{multline}
where $\phi(i,j)$ denotes the scalar potential sampled at grid point $(i,j)$, and similar for the other quantities. We can formulate a similar equation for all the various grid points, which can be used to form a square matrix system given by $[A]\{\phi\} = \{b\}$. This matrix equation can then be solved using a wide range of techniques, including specialized ones that leverage the known structure of matrices formed by FDMs. Similar approaches can be used for Maxwell's equations, which lead to the finite-difference frequency-domain (FDFD) method that has been used in various applications (for an introduction, see \cite{rumpf2014finite}).

Returning to a more general discussion about FDMs, in a 3D problem we will divide the solution space into many small rectangular cubes, which we will refer to as \textit{cells} for brevity. In this way, each cell can be assigned its own material properties (i.e., $\epsilon$, $\mu$, and $\sigma$ values). Finite difference formulas are applied at various points on the grid formed by the different cells to discretize the equations governing the physical system being studied. To accommodate this, basic FDMs treat the material properties as constants within each cell. Since most practical geometries involve many curved surfaces, this leads to what is known as a \textit{staircasing approximation} error that can greatly limit the accuracy achievable with FDM methods because the underlying geometric representation of the problem is inherently flawed. The staircasing problem can attempt to be alleviated by using more cells to better resolve the curved edges, but this makes the method more computationally expensive. Furthermore, if we naively make the grid finer throughout the entire solution space, more cells than necessary will be included in regions of the solution space that do not require a finer mesh resolution. 

\begin{figure}[t!]
	\centering
	\includegraphics[width=0.8\linewidth]{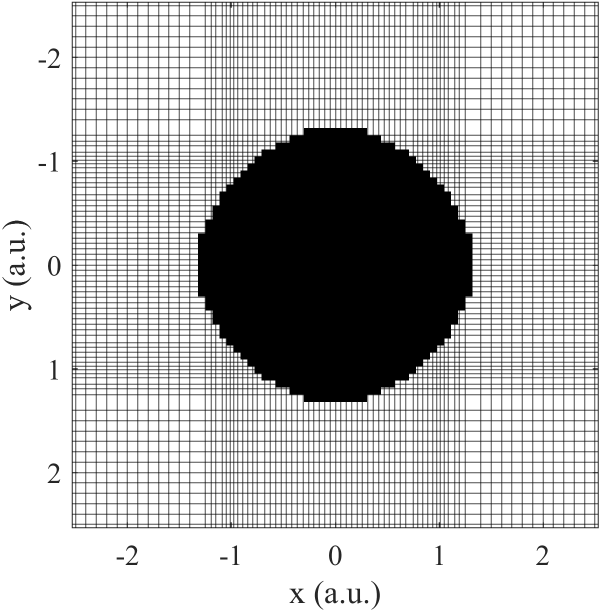}
	\caption{Example of an FDM grid using non-uniform spatial step sizes. A discretized cylinder is shown in solid black, where the staircasing approximation is apparent. The requirement for rectangular cells also forces the mesh to change along only one coordinate axis at a time, causing the ``bands'' of highly refined cells even far outside of the object of interest.}
	\label{fig:fdtd-varying-grid}
\end{figure}

This is a common problem for all numerical modeling methods, with the typical solution being local mesh refinement. In the case of FDMs, this would correspond to varying the grid spacing throughout the solution space as needed to achieve the desired geometric modeling fidelity. An example of this is shown in Fig. \ref{fig:fdtd-varying-grid}, which also highlights the challenge of this approach for FDMs. In particular, FDMs must operate on rectangular cells, so it is only possible to change the size of the mesh along one coordinate axis at a time. The effect of this is that mesh refinement in one location leads to large ``bands'' outside of the local area of the geometry that become over-discretized, as shown in Fig. \ref{fig:fdtd-varying-grid}. This lack of flexibility with mesh refinement can greatly increase the computational burden for complicated geometries, while still not fully solving the issue of the staircasing approximation.

Due to these issues, FDMs can struggle in applications with highly complicated geometries or when high accuracy is needed. It is possible to create more sophisticated FDMs that attempt to address the drawbacks discussed in this section (e.g., conformal FDMs that alleviate staircasing errors \cite{benkler2006new}), but this quickly degrades the simplicity and ease of implementation that is desirable of an FDM. As a result, for complicated scenarios and where high accuracy is needed, it can be common to move to an alternative discretization strategy like FEM. However, FDMs remain an extremely popular method due to the simplicity of their formulation and ease of implementation, and certainly are still a valuable CEM technique. For interested readers, a more detailed introduction to these methods can be found in \cite{teixeira2023finite}, while the standard reference book on these methods in CEM is \cite{taflove2005computational}.

\subsubsection{Time Domain Considerations}
\label{subsubsec:fdm-td}
In contrast to most other CEM techniques that operate in the frequency domain (i.e., they directly compute steady-state behavior), FDMs are most commonly used for time domain simulations, which are then referred to as finite-difference time-domain (FDTD) methods. The reason for this is that suitably formulated FDTD methods lead to a trivial diagonal matrix system that can be solved without resorting to using any actual matrix manipulation techniques \cite{teixeira2023finite,taflove2005computational}. This helps avoid or mitigate many of the typical challenges associated with time domain simulations; however, there do still remain important considerations for FDTD methods which we comment more on in this section.

First, it is important to note that in FDTD methods the time derivatives are also approximated with finite differences. To achieve suitable accuracy, a sufficiently small time step must be used to properly capture the temporal variation of the solution. This idea is an analog to the Nyquist sampling theory. To achieve reasonable results, the time step generally must be selected so that $\Delta t\ <\ T/20$, where $T$ is the period of the highest frequency of interest \cite{jin2011theory}. For higher accuracy, even smaller time steps are usually required.

Beyond ensuring that the time step size satisfies a reasonable number of samples per period, the time step size must also be small enough to ensure stability of the solution. A $\Delta t$ that is not sufficiently small can lead to fields with unphysical exponential growth, causing the solution to become meaningless. It can be shown that in order for the solution to be stable, the time step must satisfy
\begin{align}
    \Delta t \leq  \frac{1}{c\, \sqrt{ \displaystyle \bigg( \frac{1}{\Delta x}\bigg)^2 + \bigg( \frac{1}{\Delta y}\bigg)^2 + \bigg( \frac{1}{\Delta z}\bigg)^2   } } ,
	\label{eq:cfl-condition}
\end{align}
where $c$ is the speed of light. This condition is known as the Courant-Friedrichs-Lewy (CFL) stability condition, and places a strict bound on suitable choices for $\Delta t$ to ensure a successful simulation \cite{taflove2005computational}.

Up to this point, we have treated the required spatial step size as dependent on adequately representing the geometry of interest. However, for operating frequencies that are high enough that the wavelength is small compared to the geometry, the spatial step size must now be small enough to obtain an accurate solution. That is, the spatial step size must provide a sufficient number of samples per wavelength, usually around 20 or more (significantly more can be needed for high accuracy). In these situations, the CFL constraint is not typically a significant issue because the $\Delta t$ required for accuracy is relatively close to that needed for stability. However, the situation is significantly different if the simulation contains geometric features that are much smaller than the wavelengths of interest (these features are generally called electrically small). In these situations, it is the small required spatial step size in conjunction with the CFL condition that can quickly result in the FDTD method becoming computationally prohibitive \cite{zhen2000toward}.

For example, we can consider a coplanar waveguide geometry typical for cQED devices where a spatial step size less than 0.1 $\mu$m could easily be needed to suitably resolve the transverse geometric features. The corresponding CFL condition would result in a maximum $\Delta t$ of approximately 0.2 fs. If the thin layers in the material stackup were to be modeled as well, this spatial step size would need to be even smaller. If we were to consider an operating frequency for the simulation of roughly 5 GHz, the time step size needed from a ``Nyquist''-like accuracy perspective would be around 10 ps. Hence, for \textit{each} of the time steps needed for accuracy in simulating a particular waveform, approximately 50,000 time steps will instead have to be taken for stability purposes. Since many simulations require modeling the performance of a system over many periods of the temporal waveform, this problem can become prohibitive for even modestly-sized simulation regions. Due to these computational efficiency issues with the standard FDTD method (known as an \textit{explicit method}), alternative approaches known as \textit{implicit methods} that do not have a stability constraint have been developed.

Implicit and explicit FDTD methods differ from each other in their time-stepping dependencies. While an explicit method’s time-stepping equations depend solely on previously computed values, implicit methods have inter-dependencies in their time-stepping equations, resulting in the need to construct and solve a large linear system of equations in the form of a matrix. This leads to a method that is unconditionally stable, offering a less strict maximum time step size. These unconditionally stable FDTD methods are in principle of great interest if the accuracy of the results they produce for a substantially larger time step is acceptable. 

Unfortunately, this is found to not be the case for the primary approach in this vein of research, known as the alternating-direction implicit FDTD (ADI-FDTD) method. This is the most well-known implicit FDTD method because it yields a tridiagonal matrix system that can be efficiently solved (i.e., specialized algorithms can solve this system much faster than algorithms using general-purpose techniques). Although the time step can be arbitrarily large without loss of stability, it is found that accuracy is quickly lost by truncation error terms at a rate proportional to the square of the time increment multiplied by the spatial derivatives of the fields \cite{garcia2002accuracy}. As a result, the CFL constraint is effectively only relaxed by an order of magnitude or two at best, which is generally not sufficient to alleviate the issues discussed earlier in terms of computational efficiency. Many other implicit FDTD variants have been developed over the years, e.g., \cite{yang2011unconditionally,tan2020fundamental,grande2014accuracy,tan2007unconditionally,chen20063d,chen2007novel}, but they all essentially suffer from the same accuracy issues and so will generally struggle at modeling significantly multiscale geometries that arise frequently in designing cQED devices. We are not aware of any uses of state-of-the-art implicit FDTD methods for cQED modeling, but if combined with other advanced features (e.g., conformal discretizations) they may be able to achieve the required accuracy for these applications. 

\subsection{Finite Element Methods}
\label{subsec:fem}
We will now turn our attention to how FEM can be used in CEM analyses. Just like FDM, FEM is used to solve the PDEs that arise in various EM applications. Although similarities exist, one of the primary advantages of FEM over FDM is the improved capability for modeling complex geometries. The key to this capability is that FEM formulations approximate the solution to the PDE, while the FDTD method approximated the differential operators. It is much simpler to develop more accurate approximations to the solutions of PDEs for relatively arbitrary geometries than it is to develop higher-accuracy approximations to differential operators, so FEM formulations can utilize more realistic discretizations of complex geometries (e.g., using tetrahedral meshes) and avoid staircasing errors. Correspondingly, if the FEM analysis is performed in a suitable manner, the results can very frequently achieve excellent agreement with measured results for a fabricated device in conventional EM applications and constitutes one of the most powerful, robust, and practical numerical strategies for solving Maxwell's equations. However, unique aspects of cQED devices stress the limitations of conventional FEM approaches, which will be commented on further in Section \ref{sec:practical-considerations} after we establish the fundamental concepts.

\subsubsection{Fundamental Concepts}
\label{subsubsec:fem-fundamentals}
We will now look at the basic FEM process for taking a continuous PDE and converting it into a linear matrix equation as it informs the discussion of opportunities and challenges with these methods for practical cQED applications. Given the dominance of FEM as the method of choice for these applications, we will go into more detail on the formulation of this method. 

To make our discussion more concrete, we will consider the EM wave equation in the frequency domain given by
\begin{align}
	\nabla\times\mu_r^{-1}\nabla\times\mathbf{E} - k_0^2 \epsilon_r \mathbf{E} = -ik_0\eta_0 \mathbf{J}_\mathrm{imp},
	\label{eq:vec5}
\end{align}
where $k_0 = \omega\sqrt{\mu_0 \epsilon_0}$ is the free space wavenumber, $\eta_0 = \sqrt{\mu_0/\epsilon_0}$ is the intrinsic impedance of free space, and $\mathbf{J}_\mathrm{imp}$ is an impressed current source that produces the EM fields (many other source types are also possible). For simplicity, we will only consider one explicit boundary condition, an impedance boundary condition on the surface $S$, given by
\begin{align}
    \hat{n}\times\big( \mu_r^{-1} \nabla \times \mathbf{E} \big) + i k_0 \eta_0 Z_s^{-1} \hat{n}\times\hat{n}\times \mathbf{E} = 0,
    \label{eq:imp-bc}
\end{align}
where $Z_s$ is the prescribed impedance of the surface. This kind of condition is often used in modeling superconductors to approximately account for the kinetic inductance effect of the superconducting materials \cite{kerr1996surface,kongpop2018modeling,whitaker1988propagation}. However, alternative methods using volumetric discretizations are also possible \cite{Levenson-Falk_2025}, albeit, subject to the multiscale modeling challenges discussed in more detail in Section \ref{sec:practical-considerations}.

Now, to begin to convert (\ref{eq:vec5}) into a finite-dimensional matrix equation, we will first need to come up with a way to approximate $\mathbf{E}$ using a discrete set of variables. We do this by expanding the unknown function $\mathbf{E}$ with a set of known \textit{basis functions} (also sometimes referred to as \textit{expansion functions}) that have unknown expansion coefficients. For instance, we have
\begin{align}
	\mathbf{E}(\mathbf{r}) \approx \sum_{n=1}^N a_n \mathbf{N}_n(\mathbf{r}),
	\label{eq:fem3}
\end{align}
where $a_n$ is the unknown expansion coefficient and $\mathbf{N}_n(\mathbf{r})$ is a known continuous basis function. The exact form that $\mathbf{N}_n(\mathbf{r})$ should take is problem-specific. However, a general rule will be that $\mathbf{N}_n(\mathbf{r})$ should be able to satisfy the boundary conditions of the problem being considered and that it can make a good approximation to $\mathbf{E}(\mathbf{r})$ throughout the spatial domain of interest. Overall, the choice of $\mathbf{N}_n(\mathbf{r})$ has an incredibly important impact on the performance of a FEM formulation and must be selected carefully, as will be commented on briefly later. 

With (\ref{eq:fem3}) in hand, we now have a discrete number of variables that we need to solve for (i.e., all the $a_n$'s). If we substitute (\ref{eq:fem3}) into (\ref{eq:vec5}), we are faced with the problem that we still have an infinite-dimensional problem due to the infinite $\mathbf{r}$ values that we need to have our PDE enforced at. As a result, we will need to ``relax'' what we consider to be a solution to the problem, which is referred to as the \textit{weak solution} to the PDE. Typically, this is done by enforcing the PDE in some kind of \textit{averaged sense} by multiplying (\ref{eq:vec5}) by what is known as a \textit{testing function} or \textit{weighting function} and then integrating this over the spatial domain of interest. For a particular weighting function $\mathbf{W}_m$, we would get
\begin{multline}
	\sum_{n=1}^N a_n \! \int\! \mathbf{W}_m(\mathbf{r}) \cdot \bigg[ \nabla\times\mu_r^{-1}\nabla\times\mathbf{N}_n(\mathbf{r}) - k_0^2 \epsilon_r \mathbf{N}_n(\mathbf{r}) \bigg] dV \\ = -ik_0\eta_0  \int \mathbf{W}_m(\mathbf{r}) \cdot \mathbf{J}_\mathrm{imp}  dV.
	\label{eq:fem4}
\end{multline} 
We can repeat this process for a sufficiently large set of $\mathbf{W}_m$'s to get a system of linear algebraic equations that can be solved to find the $a_n$'s. Clearly, just as we must be careful in choosing the $\mathbf{N}_n$'s we need to be careful in how we choose the $\mathbf{W}_m$'s to ensure that our resulting matrix equation has ``good'' properties so that it can be solved numerically in an accurate way. 

To proceed, we now integrate by parts to transfer one curl onto the testing function. With a further simple application of Gauss' theorem we can convert a volume integral of the divergence into a surface integral. Combining these steps, we arrive at 
\begin{multline}
    \sum_{n=1}^N a_n \! \int \!\bigg[ \mu_r^{-1} \big( \nabla\times\mathbf{W}_m\big) \cdot\big(\nabla\times\mathbf{N}_n \big) -   k_0^2 \epsilon_r\mathbf{W}_m \cdot   \mathbf{N}_n\bigg]dV  \\  -\sum_{n=1}^N a_n\int  \hat{n}\cdot \bigg[ \mathbf{W}_m  \times \mu_r^{-1} \nabla \times \mathbf{N}_n  \bigg] dS \\ =   -ik_0\eta_0 \int \mathbf{W}_m \cdot \mathbf{J}_\mathrm{imp} dV.
	\label{eq:vec10}
\end{multline}
We can use the boundary condition (\ref{eq:imp-bc}) to rewrite the second line, yielding
\begin{multline}
    \sum_{n=1}^N a_n \! \int \!\bigg[ \mu_r^{-1} \big( \nabla\times\mathbf{W}_m\big) \cdot\big(\nabla\times\mathbf{N}_n \big) -   k_0^2 \epsilon_r\mathbf{W}_m \cdot   \mathbf{N}_n\bigg]dV  \\  -\sum_{n=1}^N a_n\int   \mathbf{W}_m \cdot \bigg[  ik_0\eta_0Z_s^{-1} \hat{n}\times\hat{n}\times \mathbf{N}_n \bigg] dS \\ =   -ik_0\eta_0 \int \mathbf{W}_m \cdot \mathbf{J}_\mathrm{imp} dV.
	\label{eq:vec11}
\end{multline}
Assembling all the equations for the different $\mathbf{W}_m$'s together, we get a matrix equation given by
\begin{align}
    \big([S] +ik_0 [Z]  - k_0^2 [M] \big)\{ a \} = \{b\},
    \label{eq:fem-fd-full}
\end{align}
where
\begin{align}
    [S]_{mn} = \int \mu_r^{-1} \big( \nabla\times\mathbf{W}_m\big) \cdot\big(\nabla\times\mathbf{N}_n \big) dV,
    \label{eq:stiffness}
\end{align}
\begin{align}
    [Z]_{mn} = \eta_0 \int   Z_s^{-1} \big( \hat{n}\times \mathbf{W}_m \big) \cdot \big( \hat{n}\times \mathbf{N}_n \big) dS,
\end{align}
\begin{align}
    [M]_{mn} = \int  \epsilon_r\mathbf{W}_m \cdot   \mathbf{N}_n  dV,
    \label{eq:mass}
\end{align}
\begin{align}
    \{b\}_m =  -ik_0\eta_0 \int \mathbf{W}_m \cdot \mathbf{J}_\mathrm{imp} dV.
\end{align}
Commonly, $[S]$ and $[M]$ are referred to as the stiffness and mass matrices, respectively, in analogy to FEM for structural analysis. While $[M]$ is typically a well-conditioned matrix, $[S]$ is very poorly conditioned due to the null space inherited from the curl-curl operators involved in the typical discretizations. This contributes to difficulties in solving problems in various operating regimes, including the multiscale structures used in cQED devices, which will be discussed more in Section \ref{sec:practical-considerations}.

With a matrix equation developed, it becomes necessary to choose the $\mathbf{N}_n$'s and $\mathbf{W}_m$'s. This can be a nuanced subject in general, but due to the maturity of this field the set of functions that one will realistically choose from is now well established \cite{jin2015finite}. The main important consideration at a high level for understanding FEM is that finding a function that satisfies the necessary properties on a global scale (i.e., over all values of $\mathbf{r}$ that are of interest in the problem) for an arbitrary 3D EM problem is essentially impossible. However, this task is significantly simpler if we make each $\mathbf{N}_n(\mathbf{r})$ only need to expand $\mathbf{E}(\mathbf{r})$ over a fairly small spatial range. This is the core idea of FEM; i.e., to expand $\mathbf{E}(\mathbf{r})$ with a set of simple functions that each have a small spatial support, but as an entire set are able to represent $\mathbf{E}(\mathbf{r})$ over all $\mathbf{r}$ to some desired level of accuracy. Once a set of $\mathbf{N}_n$'s have been selected, there are many different ways to determine a suitable set of $\mathbf{W}_m$'s. However, for most PDEs that arise in physics (including EM PDEs), an underlying symmetry can be achieved in the discretized system that often yields a ``good'' discretization of the PDE by choosing our $\mathbf{W}_m$'s to match the $\mathbf{N}_n$'s. This approach is often referred to as \textit{Galerkin's method}. 

Generally, the underlying basis functions $\mathbf{N}_n(\mathbf{r})$ will have some kind of polynomial functional form. One can construct more accurate basis functions that can lead to higher accuracy in solving an overall PDE by creating basis functions with higher polynomial orders, as will be discussed more in Section \ref{sec:practical-considerations}. This has been systematically done \cite{jin2015finite}; albeit, in most practical implementations the maximum polynomial order is not allowed to grow to significantly large orders due to diminishing returns for modeling arbitrary systems that may have very complicated geometries. 

\begin{figure}[t!]
    \centering
    \includegraphics[width=0.95\linewidth]{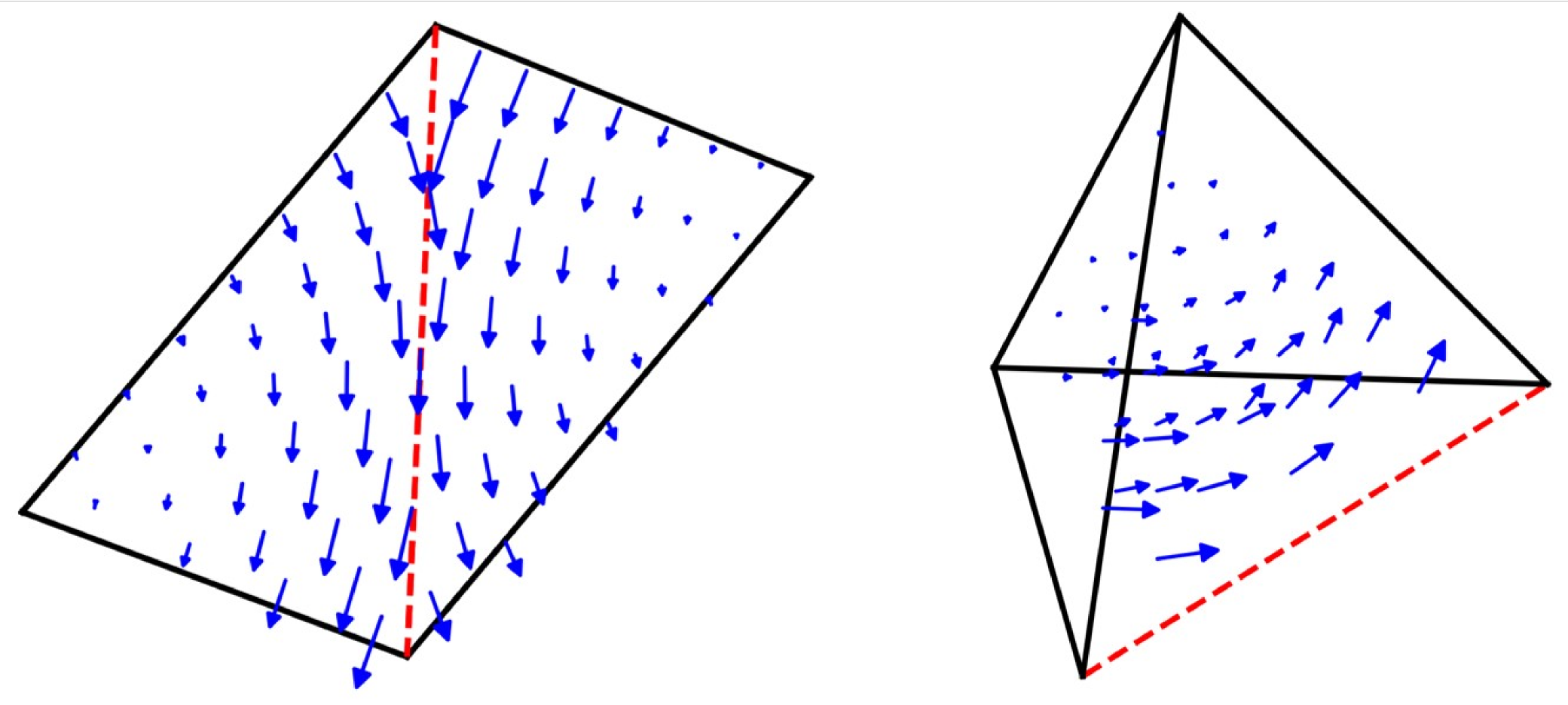}
    \caption{Example of a vector basis function, often called a vector edge element, for a 2D triangular mesh (left) and in a single tetrahedron (right). The edge the basis function is associated with is shown as a dashed red line. A critical property of this basis function is it has a constant tangential value along the edge it is associated with and is purely normal to all other edges of the mesh element. This allows the tangential continuity of fields to be exactly captured. Further, the normal components across a boundary can be discontinuous, which also preserves the correct EM boundary conditions.}
    \label{fig:edge-elements}
\end{figure}

Regarding the specific choice of basis function, for EM applications the standard approach is to use some form of vector basis function \cite{jin2015finite}. In their simplest form, these functions are associated with each edge of the computational mesh, and have been derived for a wide range of mesh types; albeit, tetrahedral meshes are the most commonly used. Importantly, these vector basis functions have been carefully designed so that they can faithfully capture key aspects of EM fields. For example, they ensure tangential continuity of the fields between different mesh elements, preserving a key EM boundary condition. An example of a vector basis function is shown in Fig. \ref{fig:edge-elements} for a triangular mesh and in a tetrahedral element. Prior to the adoption of these vector basis functions, FEM analysis in CEM was plagued with poor numerical behavior, supported spurious (i.e., non-physical) solution modes, among other issues \cite{jin2015finite}. All these issues were resolved through the use of vector basis functions that correctly preserved fundamental properties of the EM fields, such as the boundary conditions between different regions. 

\subsubsection{Time Domain Considerations}
\label{subsubsec:fem-td}
While most FEM analysis is done in the frequency domain, these techniques can still be used in the time domain. In this situation, they are often referred to as finite element time domain (FETD) methods. A similar spatial discretization process can be used in this case; however, the temporal discretization must be handled differently. If we were to consider directly computing the inverse Fourier transform of (\ref{eq:fem-fd-full}), we would find that depending on the nature of the surface impedance $Z_s$ certain complications could arise. In a general-purpose tool, no assumptions of $Z_s$ would typically be made, requiring the user to either have it be purely resistive or resort to complicated treatments to handle more arbitrary frequency dependence (e.g., the method of \cite{2015_Sharbaf_Dispersion}). However, if we restrict ourselves to impedances that are capacitive or inductive, we find that the resulting terms could easily be ``lumped'' into the definitions of $[M]$ or $[S]$ due to the common temporal dependence. Although this is in practice simple to do, not every commercial or open-source tool may be implemented to handle this. 

For simplicity, in this section we will ignore surface impedances and just consider the time domain wave equation given by
\begin{align}
    [M] \frac{d^2}{dt^2} \{a\}  + [S]  \{a\}  = \{b\}.
	\label{eq:fetd7}
\end{align}
There are various ways to now discretize the temporal aspects of this equation, but the most common and simplest is to use a finite difference approximation. If we use central differences for this, we arrive at a time-stepping equation of
\begin{multline}
      [M] \{ a \}^{n+1} = \bigg\{ 2 [M] - \Delta t^2 [S] \bigg\} \{ a \}^{n}  \\ -  [M]  \{ a \}^{n-1} + \Delta t^2 \{b\}^n,
	\label{eq:fetd12}
\end{multline}
where a superscript $n$ denotes which time step the vector is sampled at. Clearly, we can use this equation to advance our solution in time given proper initial conditions for $\{a\}$.

One main difference between (\ref{eq:fetd12}) and the explicit time-marching methods that are common with FDTD methods is that because $[M]$ is a non-diagonal matrix we will need to solve a matrix equation during each time step of our simulation. For practical problems with large matrix sizes, an iterative solver must then be used in every time step. To make this process efficient, we may want to form an effective preconditioner that we can use to improve the convergence of the iterative solver. Since the matrices do not change throughout the simulation, we only need to form this preconditioner once. 

While this has some appealing properties compared to frequency domain methods, the reality is that it typically takes much longer to perform one time domain simulation than it does to perform one frequency domain simulation. With the advent of efficient interpolating frequency sweep techniques (which are implemented in commercial tools) \cite{jin2015finite}, one typically only needs to run a relatively small subset of frequency points in a given frequency band to characterize a device, making this approach typically more efficient than a time domain method while still achieving suitable accuracy for standard microwave engineering applications. Where time domain methods can begin to become of significant interest is for applications where frequency domain simulations are no longer suitable. This can be the case for certain kinds of nonlinear modeling problems and in multiphysics modeling applications where the different physics involved are tightly coupled \cite{jin2019multiphysics,wang2008symmetric,yan2018nonlinear,chen2020multiphysics,ramachandran2023charge,roth2024maxwell,2024_Elkin_JTWPAs,elkin2025ims}. There could also be advantages for finding large numbers of eigenpairs, as will be discussed in Section \ref{sec:eig-overview}.

Returning to the temporal discretization, similar to FDTD there are also implications to the \textit{stability} of the method. A stability analysis shows that the central difference formulas used here results in a conditionally stable method. However, because the FETD approach has an unstructured grid we are no longer able to derive a ``simple'' stability condition in the way that is possible for FDTD. Instead, the stability condition can only be derived in terms of the properties of the matrices involved for a particular problem, see \cite{jin2011theory,jin2015finite,jiao2002general}. However, there is an approximate result that the stability condition can usually be estimated as
\begin{align}
	\Delta t < 0.3 h_\mathrm{min} / c
	\label{eq:fetd13}
\end{align}
for first-order finite elements, where $h_\mathrm{min}$ is the size of the smallest element in the mesh and $c$ is the relevant speed of light \cite{jin2015finite}. This estimate can be extended to higher-order elements if $h_\mathrm{min}$ is divided by the order of the element used before plugging into  (\ref{eq:fetd13}).

For applications where this stability constraint is restrictive, it is possible to formulate an unconditionally-stable time marching system by using the Newmark-$\beta$ integration method \cite{jin2015finite}. This approach is equivalent to using central differencing for first and second order derivatives and using a specially-designed weighted average for the undifferentiated quantities. The most common choice for this results in a method which is both unconditionally stable and second-order accurate. For this method, the time-stepping formula becomes 
\begin{multline}
	\bigg\{ \frac{1}{(\Delta t)^2} [M] + \beta [S] \bigg\} \{ a \}^{n+1} = \bigg\{ \frac{2}{(\Delta t)^2} [M] \\ - (1-2\beta)[S] \bigg\} \{ a \}^{n}  - \bigg\{ \frac{1}{(\Delta t)^2} [M] +\beta[S] \bigg\} \{ a \}^{n-1} \\ + \beta \{f\}^{n+1} + (1-2\beta)\{f\}^n + \beta \{f\}^{n-1},
	\label{eq:fetd14}
\end{multline}
where $\beta \geq 1/4$ for unconditional stability. Although there are a number of differences between the two time-stepping formulas given in (\ref{eq:fetd12}) and (\ref{eq:fetd14}), the main one of interest is that the matrix $[S]$ occurs in the left-hand side of (\ref{eq:fetd14}) but does not in (\ref{eq:fetd12}). This leads to ill-conditioning for certain situations, including for multiscale structures, which degrades the convergence of an iterative solver compared to one used for (\ref{eq:fetd12}). Despite this challenge, these unconditionally-stable FETD approaches can still be made practical, and importantly only require temporal sampling choices to be made purely on considerations of temporal sampling accuracy \cite{white2004full,wang2010application}, which is in contrast to the implicit FDTD methods discussed in Section \ref{subsubsec:fdm-td}. 

\subsection{Method of Moments}
\label{subsec:mom}
The final ``main'' CEM technique is MoM, which is also sometimes referred to as the \textit{moment method} or the \textit{boundary element method (BEM)}. At a high level, MoM follows essentially the same process as FEM but is used to discretize integral equations rather than PDEs. Although this may seem like a ``small change'', the formulation and solution of integral equations is so different from working with PDEs that the MoM is well and truly a distinct computational technique to be considered.

\subsubsection{Fundamental Concepts}
\label{subsubsec:mom-fundamentals}
To gain some basic insight into MoM, we will discuss its application to the electrostatics problem of computing the capacitance of a perfect electric conductor (PEC) structure embedded in a homogeneous medium. Detailed derivations of integral equations for the EM wave equation can be found in various references, such as \cite{jin2011theory,chew1995waves}, but are omitted here due to their more substantial complexity. To begin, we will note that EM integral equations are generally cast in the form of some kind of surface source being integrated against a Green's function for a related problem. For our electrostatics problem of interest, the differential equation that the integral equation will solve is Poisson's equation, which in 3D is
\begin{align}
	\nabla^2\phi(\mathbf{r}) = -\rho/\epsilon.
	\label{eq:intro1}
\end{align}
We can also define a Green's function for this problem that satisfies (\ref{eq:intro1}) for a point source excitation; namely, 
\begin{align}
	\nabla^2 g(\mathbf{r},\mathbf{r}') = -\delta(\mathbf{r}-\mathbf{r}'),
	\label{eq:intro2}
\end{align}
where $g(\mathbf{r},\mathbf{r}')$ is the Green's function. Using standard methods, it is possible to determine that the solution to (\ref{eq:intro2}) is
\begin{align}
	g(\mathbf{r},\mathbf{r}') = \frac{1}{4\pi \abs{\mathbf{r}-\mathbf{r}'}}.
	\label{eq:static-green}
\end{align}

We can now derive our integral equation by using the Green's function to invert Poisson's equation. We start by integrating Poisson's equation against the Green's function to get
\begin{align}
	\iiint g(\mathbf{r},\mathbf{r}') \nabla^{'2}\phi(\mathbf{r}') \, dV' = -\iiint g(\mathbf{r},\mathbf{r}') \frac{\rho(\mathbf{r}')}{\epsilon} \, dV' ,
\label{eq:green1}
\end{align}
where $dV'$ means we are integrating over the primed variables in the volume $V$. We now transfer the derivatives from $\phi$ onto $g(\mathbf{r},\mathbf{r}')$ in (\ref{eq:green1}) using integration by parts repeatedly. This gives us
\begin{align}
       \iiint \big[ \nabla^{'2} g(\mathbf{r},\mathbf{r}') \big] \phi(\mathbf{r}') \, dV' = -\iiint g(\mathbf{r},\mathbf{r}') \frac{\rho(\mathbf{r}')}{\epsilon} \, dV' ,
	\label{eq:intro3}
\end{align}
where we have ignored the boundary terms from the integration by parts because they are located at infinity where these functions go to 0. We can now make use of the fact that derivatives with respect to primed and unprimed variables only differ by a minus sign for the Green's function here so that
\begin{align}
	\iiint \big[ \nabla^{2} g(\mathbf{r},\mathbf{r}') \big] \phi(\mathbf{r}') \, dV' = -\iiint g(\mathbf{r},\mathbf{r}') \frac{\rho(\mathbf{r}')}{\epsilon} \, dV' .
\end{align}
Our final step is to use the definition of the Green's function via the differential equation it satisfies given in (\ref{eq:intro2}) to substitute in the Dirac delta. Using its sifting property, we arrive at Coulomb's law as
\begin{align}
	\iiint g(\mathbf{r},\mathbf{r}') \frac{\rho(\mathbf{r}')}{\epsilon} \, dV' = \phi(\mathbf{r}),
    \label{eq:coulomb1}
\end{align}
which we can now turn to solving using MoM.

We begin by assuming that we have a perfect conductor that has some unknown surface charge distribution. The potential produced at an observation point $\mathbf{r}$ by this surface charge distribution can be determined from (\ref{eq:coulomb1}) where we now only need to integrate over the surface $S$ of the metallic object where the charge density is located. Since $\phi(\mathbf{r})$ is not known at general $\mathbf{r}$ and $\rho$ is also not known over $S$, to proceed we need to take our observation point $\mathbf{r}$ to be on the surface of the conductor where we know that the potential must be a known constant value, denoted as $\phi_0$, that comes from the boundary condition of the problem. In this case, our integral equation becomes
\begin{align}
	\iint g(\mathbf{r},\mathbf{r}') \frac{\rho(\mathbf{r}')}{\epsilon} dS' = 	\phi_0, \,\,\,\, \mathbf{r} \in S.
	\label{eq:integral1}
\end{align} 

We can now go about solving this integral equation for $\rho$ by following the same basic process as FEM. Namely, we can subdivide the surface $S$ up into smaller portions where we can express $\rho$ using simple basis functions, denoted by $v_n$. While different surface meshes are possible, a triangular one is most commonly used. We can then test the integral equation by multiplying by a testing function $w_m$ and then integrating over the surface. This results in a matrix equation given by $[A]\{x\} = \{b\}$, where
\begin{align}
	[A]_{mn} = \iint_S \iint_S w_m(\mathbf{r})  \frac{g(\mathbf{r},\mathbf{r}')}{\epsilon} v_n(\mathbf{r}') dS' dS,
	\label{eq:sys-mat}
\end{align} 
\begin{align}
	\{b\}_m = \iint_S w_m(\mathbf{r}) \phi_0  dS.
\end{align}
We can solve this matrix equation to recover the surface charge density. With the surface charge density, we can then numerically evaluate Coulomb's law to determine the potential at any point in space. We can also compute the total surface charge and divide by the applied potential $\phi_0$ to compute the capacitance as $C = Q/\phi_0$, where $Q$ is the total surface charge.

Of course, this all depends on choosing suitable functions for $v_n$ and $w_m$. In contrast to FEM, the space of functions that a suitable MoM discretization can be achieved with is rather large. For instance, because we do not have to evaluate any derivatives of our basis or testing functions in (\ref{eq:sys-mat}), we can use basis or testing functions with lower order than the first-order functions that were suitable for FEM. The choice of testing functions is also expanded compared to FEM. Although relatively simple basis and testing approaches can be made to work acceptably for some electrostatic applications, this is not the case for full-wave EM integral equations. In these cases, basis and testing functions similar to those from FEM can be utilized, with some modifications to make them suitable for interpolating the surface current densities that are the unknown to be solved for in those equations \cite{rao1982electromagnetic,jin2011theory}.

Now, if $m$ and $ n$ are such that $w_m$ and $v_n$ are far apart from one another spatially, we can use simple integration methods to evaluate the integrals. However, when this is not the case, we must be careful in how we handle the singularity of the Green's function, including when $\mathbf{r}'$ may equal $\mathbf{r}$. Ultimately, these problems have been studied for decades, so mature approaches exist for handling these singular integrations (or near singular integrations) for typical basis and testing functions. Although these methods are not ``perfect'', they are generally robust enough for most use cases encountered in practice.

\subsubsection{Comparison with the Finite Element Method}
Now, from the earlier discussion we can see that the basic discretization process of MoM exactly matches the approach that was discussed in the context of FEM. Although these similarities exist, there are many important differences between FEM and MoM due to the differences in solving differential and integral equations. Some of the main differences are the following.

When solving differential equations, we discretize the entire volume of interest. Integral equations provide us with an approach that only requires discretizing the surfaces between regions with different material properties. For particularly large problems, this reduction in dimensionality can be quite important to keep the size of the matrix equation smaller.
	
When solving differential equations, we typically must determine artificial boundary conditions to terminate open region problems (e.g., absorbing boundary conditions or perfectly matched layers \cite{jin2011theory,jin2015finite}). These approximate boundary conditions contribute to numerical error. In contrast to this, integral equations do not require any artificial boundary because the Green's function used in the formulation automatically encodes the correct behavior into the solutions. 
	
For FEM, the system matrix was extremely sparse. For MoM, the system matrix is completely dense; i.e., every element in the matrix is nominally non-zero. The reason for this is that the Green's function is able to link every single basis and testing function to one another. The fact that MoM produces dense matrices greatly changes the numerical linear algebra solution approaches that should be used when solving MoM matrix equations versus FEM matrix equations. However, because typical EM Green's function do exhibit a $1/|\mathbf{r}-\mathbf{r}'|$ dependence, we can typically conclude that the MoM matrices will be diagonally dominant and that for sufficiently far distances the interactions between basis and testing functions can become ``simplified''. It is this aspect that gets exploited to design fast algorithms that allow integral equation methods to compete and, in many cases, outperform FEM for very large system analysis, as will be discussed in Section \ref{subsubsec:mom-fast}.
	
For FEM, the integrals needed to evaluate matrix elements can often be done analytically or using fairly simple numerical integration. In MoM, the singularity in the Green's function makes both the analytical and numerical evaluation of these integrals much harder. However, because to formulate these integral equation methods we must know how to effectively deal with these Green's function singularities (e.g., see \cite{wilton1984potential,yla2003calculation}), it does mean that integral equations may be able to compute fields close to conducting surfaces more accurately than FEM. This could be valuable for improving the accuracy of computing surface participation ratios that are often used to predict the influence of dielectric loss on qubit coherence for a prospective design (e.g., see the methodologies discussed in \cite{wenner2011surface,wang2015surface,gambetta2016investigating}). However, it is still important to recognize that near field singularities (e.g., next to conducting corners or wedges) the surface currents will be singular as well \cite{jin2011theory}, which can be a dominant error source in numerical solutions when using standard basis functions to discretize the surface currents \cite{warnick2008numerical}. Specialized basis functions to try and account for these singularities have been researched (e.g., see \cite{peterson2016basis,graglia2018computation}), but these techniques are not in widespread use and complicate the implementation of any code. Regardless, testing the use of such methods for evaluating surface participation ratios could be valuable for qubit design.

\begin{figure}[t!]
    \centering
    \includegraphics[width=0.7\linewidth]{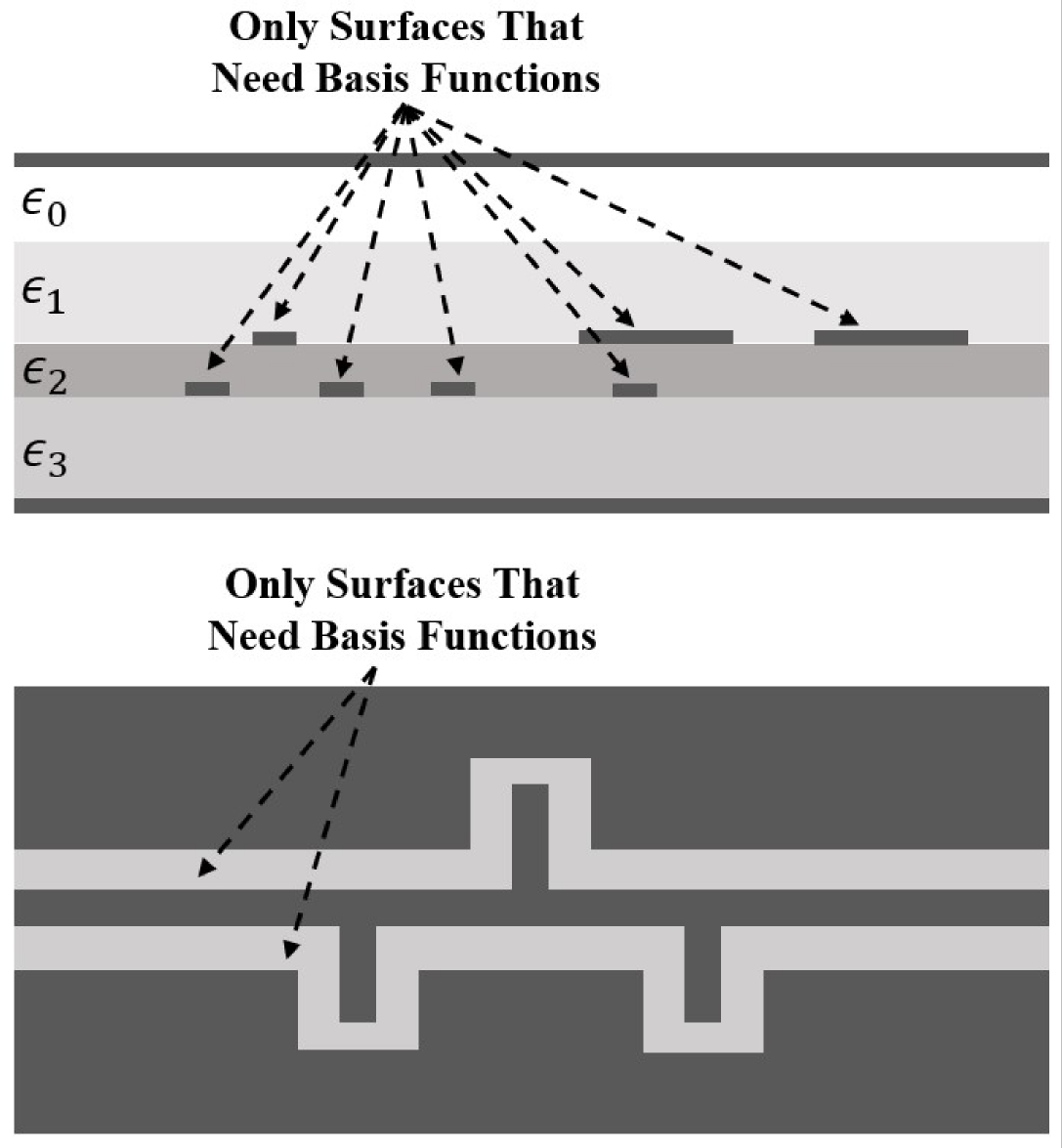}
    \caption{Examples of how specialized Green's functions can reduce the number of unknowns for different problems. (Top) A cross-sectional view of a multi-layer circuit where a Green's function for layered media can be used so that only the circuit traces require basis functions. (Bottom) A top-view of a coplanar waveguide structure where a different specialized Green's function can be used so that only the gaps between conductors require basis functions.}
    \label{fig:special-integral-equations}
\end{figure}

Another fundamental difference between differential and integral equations is the amount of specialization that can be done with integral equations. In particular, specialized Green's functions and corresponding integral equations can be formulated for different classes of geometries to greatly modify the computational workload of the algorithm, as illustrated in Fig. \ref{fig:special-integral-equations}. For instance, integral equations can be formulated using Green's functions for layered media (in open or closed regions) so that the only locations where basis functions are needed are on structures that are different from the background layered medium \cite{michalski2002multilayered,okhmatovski2024theory}. Although there are various uses for such methods, they are particularly relevant for modeling multi-layer circuit boards or chips where basis functions are now only needed on the routed circuit traces rather than on every material interface. Similar methods can also be formulated for coplanar waveguide-like structures so that basis functions are only needed along the gaps between the conductors \cite{drissi1991analysis,dib1991theoretical,wu1995full}. Although these specializations can vastly reduce the matrix dimension of a problem, it will generally be more complicated and time-consuming to evaluate each of the matrix entries. Further, the formulation and implementation of these methods are labor intensive and, by their nature, are not applicable to more general cases if the device to be modeled begins to deviate from the assumptions used in formulating the integral equation. 

\subsubsection{Fast Algorithms}
\label{subsubsec:mom-fast}
As mentioned earlier, one of the main differences between FEM and MoM is that MoM matrices are completely dense. As a result, matrix-vector products require $O(N^2)$ operations, which can become prohibitive for iterative solvers as the matrix size grows, as was shown in Fig. \ref{fig:computational-scaline}. For many years, this computational bottleneck appeared to severely limit the feasibility of using MoM for practical engineering analyses, even accounting for them only using surface discretizations. These issues were eventually addressed through the creation of \textit{fast algorithms}, which in the context of CEM almost always refers to a method used to speed up the solution of an integral equation solver \cite{chew2001fast}. Originally, fast algorithms were only applicable to speeding up matrix-vector products and so could only accelerate iterative solvers. However, more recently, new fast algorithms have been developed to accelerate direct solvers. With the advent of fast algorithms, the computational complexity of integral equation solvers was able to reach $O(N\log N)$, greatly increasing the applicability and popularity of integral equation solvers. This has made them the standard approach for analyzing electrically-large problems when the accuracy of a full-wave solution is needed. Although this is still generally the case, it should be noted that advanced techniques applied to differential equation solvers can still allow them to be applied to very large-scale problems as well (e.g., domain decomposition methods that will be discussed in Section \ref{subsec:domain-decomposition}). As a result, all of these methods continue to be areas of active research interest. 

The specific formulation details of the various fast algorithms are too involved to discuss here, but basic introductions to the two ``standard'' approaches that are most often encountered in commercial tools today can be found in, e.g., \cite{jin2011theory}. These two methods are the \textit{adaptive cross approximation (ACA)} \cite{zhao2005adaptive} and the \textit{multilevel fast multipole aglorithm (MLFMA)} \cite{song1997multilevel}. Both of these algorithms attempt to exploit the fact that as groups of sources (i.e., basis functions) and receivers (i.e., testing functions) become further apart, they tend to ``feel'' the net effect of the group of sources in a simpler way. At a coarse level, one can think of this from the perspectives of an antenna array, where close in to the array the fields received at some test point from each array element can be dramatically different, but as one moves very far away from the array the only wave that can effectively be seen is a simple plane wave. 

In the context of numerical linear algebra, this effect leads to many of the off-diagonal blocks of the MoM matrix being \textit{low rank}. Low-rank matrices can be compressed in many ways, allowing for linear algebra operations to be performed on them using dramatically less storage and processing time than just directly using all the elements of the matrix in a naive way. In the case of ACA, an algebraic approach is utilized to directly calculate this compressed form of the low rank matrix subblocks without needing to ever fully compute the subblocks. In the case of MLFMA, the factorization and compression is done based on the physics of the problem by mathematically reformulating the method. 

The main tradeoff between these approaches is that the algebraic approach of ACA will only approximately exploit the ``low rank'' nature of the physics, leading to somewhat worse performance in terms of asymptotic scaling than the MLFMA. However, the physics-based nature of MLFMA means that any time a new problem is formulated that uses a different Green's function (e.g., for layered media or other specialized integral equation methods) a new version of the MLFMA factorization must be devised and implemented, which can be significantly labor intensive. The ACA can meanwhile be applied to any MoM matrix so long as one is able to compute individual elements of the matrix when requested. Other more advanced algebraic-based fast algorithms do exist and are being actively researched (such as hierarchical matrix methods \cite{borm2003introduction} or ones using tensor trains \cite{chen2019sparsity}), but these share the same basic properties of the ACA as laid out here, so we will not discuss them in more detail.

\subsubsection{Time Domain Considerations}
\label{subsubsec:mom-td}
The time domain version of MoM is typically referred to as a time domain integral equation (TDIE) method. Historically, TDIEs have never gained wide popularity due to their significant complexity in implementation and the extreme challenges in achieving stable discretizations, despite decades of research on the subject \cite{ha2003retarded,weile2004novel,wang2008finite,shanker2009time,pray2012stability,van2016stability,roth2018development,roth2020stability}. Although workable strategies have generally been developed for perfect conducting geometries, extending these same approaches to be stable for penetrable scatterers remains a challenge (e.g., \cite{roth2021lorenz}). On top of these challenges, the TDIEs inherit the other complications of integral equation methods in needing to deal with singular integrals and different numerical breakdowns (e.g., due to multiscale discretizations). Further, the fast algorithm counterpart to the MLFMA, the plane-wave time domain (PWTD) method \cite{shanker2003fast}, has its own stability and breakdown problems beyond the already challenging ones of the underlying TDIE. Beyond this, the efficiency of the PWTD methods have never been able to seriously rival those of MLFMA for basic spectrum analysis applications. Hence, although TDIEs have some potentially interesting properties, the severe challenges involved have held them back from being widely utilized in practical engineering applications and we are not aware of any commercial tools using this approach.

\subsection{Other CEM Methods}
\label{subsec:cem-other}
While FDM, FEM, and MoM make up the primary strategies for discretizing EM equations, there are a number of other methods and terminologies that may be encountered in various contexts. Here, we attempt to provide basic details on these concepts to help disambiguate the bevy of computational algorithms that have been developed over the years and provide context on whether these methods are likely to be useful for cQED modeling applications or not.

\subsubsection{Asymptotic Methods}
\label{subsubsec:asymptotic-driven}
In typical CEM jargon, an \textit{asymptotic method} most frequently refers to a technique targeted towards especially ``high frequency'' problems where some of the wave physics effects can be well approximated with ray tracing concepts \cite{jin2011theory}. Historically, these methods were built in stages under various optics approximations. More recently, the basic ideas of these methods have been combined to create the \textit{shooting and bouncing ray method} \cite{ling1989shooting}. Ultimately, these methods attempt to decompose a wavefront into a set of rays that will be ``traced'' throughout a geometry. When these rays hit a surface in the geometry, they are reflected according to the rules of geometric optics (e.g., following Snell's laws) which requires the underlying assumption that the surface the ray is reflecting off of is very large and flat relative to the wavelength. Due to this, asymptotic methods are typically used to calculate the EM scattering from objects that are hundreds of wavelengths large (or even larger). Overall, these approaches are not relevant for multiscale systems or those that contain important resonant geometric features, making them inapplicable to the modeling of superconducting circuit quantum devices. 

Although not always categorized as ``asymptotic methods'', another class of solvers in this broad category are quasi-static methods. These are ``asymptotic'' in the sense that certain frequency-dependent effects are assumed limited enough that approximations can be made to simplify the numerical method. For instance, in certain integral equation methods this may involve using a static Green's function rather than dynamic ones, or for differential equations one may solve Poisson's equation rather than the wave equation. Generally, these quasi-static methods are used to extract lumped circuit representations of 3D structures, which is a common modeling step for many cQED designers \cite{Levenson-Falk_2025,shanto2024squadds}. Regarding actual discretization strategies, since these quasi-static methods are still integral or differential equations they can be discretized with FDM, FEM, or MoM (or the other methods discussed in Section \ref{subsec:cem-other}). When the underlying approximations are valid, solving these quasi-static equations can greatly simplify the numerical method and lead to improved solver speed. 

Unfortunately, one fundamental problem with quasi-static approaches is that it is application-dependent and often difficult to determine when these tools can be accurate for an analysis. Further, these quasi-static methods do not smoothly interpolate the behavior from static solvers to dynamic solvers \cite{zhu2011rigorous}, so there will always be some inherent error when ``passing'' between solvers in a multi-stage analysis. As an example, this can cause a number of issues in signal integrity modeling applications where a consistent set of data over a wide range of frequencies is desired to help infer the time domain response of a device to fast clock signals. As a result, modern CEM research has focused more on developing dynamic solvers that can be applied accurately and efficiently across the whole frequency spectrum of interest. This requires a method that avoids the ``low frequency breakdown'' or other challenges associated with multiscale geometries, and remains an area of active research.

\subsubsection{Hybrid Methods}
\label{subsubsec:hybrid-methods}
Hybrid methods attempt to blend distinct methods together (e.g., using FDM in one part of the system and FEM in another) \cite{jin2011theory}. There are too many variations that have been studied over the years to list them all exhaustively, but the main hybrid method that is available in commercial tools is the finite element boundary integral (FEBI) method \cite{jin2015finite}. This approach combines FEM and MoM to address the challenges in accurately terminating an FEM simulation domain for open problems (e.g., radiation and scattering problems). The idea is to perform FEM in the ``interior region'' where the geometry is highly inhomogeneous and complex (something that FEM can often do well at in comparison to MoM), and then to perform a ``boundary integral'' on the outside surface of the ``interior region'' that will be solved with MoM. This boundary integral accurately accounts for how the fields will radiate, propagate, or leak away from the interior region, and so can be placed quite close to the device being modeled in the interior region. This is in contrast to other approximate terminating conditions for the FEM domain like absorbing boundary conditions and perfectly matched layers which must be placed far enough away from the device being modeled so that the fields reaching the outer boundary are already ``plane wave-like'' \cite{jin2015finite}. However, because most cQED modeling problems are concerned with a device in a relatively closed package, FEBI is not likely to be needed for most cQED simulations. 

\subsubsection{Discontinuous Galerkin Time Domain Method}
\label{subsubsec:dgtd}
Another approach of note is the discontinuous Galerkin time domain (DGTD) method. This approach utilizes some FEM-like approaches in the discretization of the numerical system, but does this in such a way that enables a more parallelizable numerical implementation than standard FEM \cite{chen2012discontinuous,jin2015finite}. The method is also highly flexible, allowing for tailored discretizations and time-stepping schemes in different regions of the system \cite{zhang2021arbitrary,jia2023time}. However, this high flexibility does increase the complexity in implementation procedures and also leads to many variations of the method existing, all with their own unique challenges that must be tackled when being applied to a particular application at hand. For example, some variants will not properly resolve certain energy conservation laws \cite{jin2015finite}. One other point of note is that as the name suggests, this method is for time domain implementations and hence must deal with stability and time-stepping considerations as well. Overall, DGTD is a powerful technique that can find utility in many modeling applications, although we are not aware of researchers having tried to use it for cQED applications yet. 

\subsubsection{Spectral Element Method}
\label{subsubsec:sem}
The spectral element method (SEM) is another methodology that leverages similar concepts to FEM \cite{lee20063,lee20073}. There are various forms of SEMs used in EM modeling, including ``pseudo-spectral element methods'', but for simplicity we will refer to all of these as SEMs here. The main concept behind SEM is to use specialized higher-order basis functions combined with unique numerical integration routines to create special forms of matrix equations that can then be solved advantageously. Due to the use of significantly high-order basis functions, these methods also typically need to use very high-order curvilinear mesh elements which can be difficult to generate for complicated geometries. The combination of these techniques leads to a method that in principle exhibits very impressive convergence properties in the accuracy of the solution. However, these impressive accuracy results have primarily been demonstrated on very simple geometries with special meshing procedures designed for the particular geometry being modeled, and usually require the underlying solution to be ``smooth enough'' to benefit from these higher-order approximations. In many typical modeling scenarios, these conditions are not often met, so the use of SEMs in practical engineering analysis has been limited to date, and we are not aware of commercial tools utilizing this CEM approach.

\subsubsection{Discrete Exterior Calculus Method}
\label{subsubsec:dec}
Another set of hybrid-like techniques are numerical methods implemented using the formalism of discrete exterior calculus (DEC) \cite{desbrun2005discrete}. DEC is a theory formulated entirely in a discretized setting that is meant to replicate many of the properties of exterior calculus that is often used in differential geometry and for physics in curved manifolds \cite{stone2009mathematics}. The main benefits of this approach is that it helps ensure the discretization maintains important aspects of the physical theory (e.g., conservation laws) \cite{moon2015exact,mohamed2016discrete}. Although this is certainly an appealing property, most major numerical methods already determined suitable fixes to the kinds of problems DEC is meant to fix by the time DEC was being developed, limiting its use cases. One area where DEC could be advantageous is in multiphysics modeling scenarios where it can provide a uniform ``language'' to use in forming physically-conforming discretizations that will avoid spurious behavior due to unintentional violations of certain physical symmetries or conservation properties in the discrete case. 

Regarding its actual implementation, at a high level, DEC methods are similar to FDMs that leverage some concepts from FEM to be able to operate on an unstructured mesh (e.g., made with tetrahedral elements). As a result, these methods avoid the staircasing errors that FDMs suffer from. However, due to the way that the FEM concept is leveraged in these methods, it is not very achievable to form higher-order discretizations, limiting the convergence properties of the DEC method in practical scenarios. This is a significant constraint for cQED modeling applications, as discussed in more detail in Section \ref{subsec:higher-order-basis}, and so this method will likely struggle to compete with other modeling approaches for this application space. For examples of DEC methods, traditional EM modeling applications are considered in \cite{chen2017electromagnetic,zhang2022phi} while the electrodynamics of superconductors is considered in \cite{pham2023flux}. 

\subsubsection{Partial-Element Equivalent-Circuit Method}
\label{subsubsec:peec}
Integral equations in EM can be formulated and discretized in many different ways to create techniques specialized to particular modeling scenarios. One prominent ``special'' integral equation formulation is the partial-element equivalent-circuit (PEEC) method \cite{antonini2023partial,ruehli2017circuit}, which has traditionally been developed and optimized for EM compatibility and signal integrity analysis applications in complicated integrated circuits (e.g., in a computer processor). The basic concept of the PEEC method is to interpret the ``electric field integral equation'' \cite{jin2011theory} in terms of Kirchoff's circuit laws to create equivalent circuit models of complicated integrated circuit structures that can then be seamlessly incorporated with other circuit models (e.g., lumped circuit models of different components like resistors, capacitors, transistors, etc.). This allows a single circuit model of a device to be constructed that includes many non-ideal effects (e.g., layout parasitics) that can then be solved using standard circuit solving techniques like modified nodal analysis (MNA). 

To improve its simplicity and performance for the intended applications, PEEC methods focus on using meshes that are well-suited to the straight traces used in many integrated circuits. In typical MoM, a triangular mesh is used so that arbitrary geometries can be represented well, but this complicates the implementation of the integral equation method. PEEC typically uses rectangular mesh cells (or quadrilateral or hexahedral mesh cells for more sophisticated implementations) for which simpler basis and testing functions are also used. These simpler basis and testing functions also enable simplified formulas to sometimes be used in evaluating the various integrals involved in the formulation, again simplifying the method implementation in comparison to a more standard MoM solver. 

Overall, for the geometries that PEEC was designed for, one can expect to need to use fewer mesh cells than a standard MoM approach. However, comparing accuracy between the methods for a fixed matrix size is more difficult since PEEC methods have traditionally been optimized for applications where accuracy requirements were relatively low; however, with a sufficiently dense mesh the PEEC method can certainly achieve accurate results. Depending on how the PEEC method is implemented, the kind of post-processing results that can be made available to the user may also be different from a standard MoM approach. For instance, computing the fields at arbitrary points in the system geometry will almost certainly be possible with any standard MoM approach, but this functionality may not be implemented in a PEEC method that is more concerned typically with monitoring the voltages and currents at various nodes of a circuit.

Although PEEC methods provide flexibility in being optimized for particular applications, they still can quickly lead to the production of large MNA matrices that must be solved numerically. Fundamentally, this is still done using direct or iterative solvers, as discussed in Section \ref{subsec:direct-vs-iterative}. The solution of these equations can be sped up using similar fast algorithms as those used in standard MoM approaches, which were discussed in Section \ref{subsubsec:mom-fast}.

Overall, PEEC methods optimized for cQED applications could have interesting properties and be a competitive simulation approach. However, careful testing of this would still be required given the many specializations and approximations that can often be utilized in formulating a PEEC method. It is also worth emphasizing that the term ``PEEC'' is often used to describe a broad class of implementations for which many different variations, approximations, and simplifications have been studied. Hence, not all PEEC solvers will necessarily have comparable properties; and so, the conclusions made from using one PEEC solver may not be applicable to a different PEEC solver.

\subsection{Domain Decomposition Methods}
\label{subsec:domain-decomposition}
As the dimension of the matrix system to be solved continues to grow, even iterative solvers can become challenging to use efficiently. This is particularly so for the more general Krylov subspace-based iterative solvers that require storing and manipulating previous vectors from the Krylov subspace. Generally, the convergence of the method will be improved as more of the Krylov vectors are retained. However, for large system sizes, the storage of these vectors can become a bottleneck on parallel architectures, requiring the algorithm to be ``restarted'' more often by forming a new Krylov subspace \cite{saad1986gmres}. For particularly challenging modeling problems, this may prevent the iterative solver from reaching convergence in a feasible amount of time.

\begin{figure}
    \centering
    \includegraphics[width=0.95\linewidth]{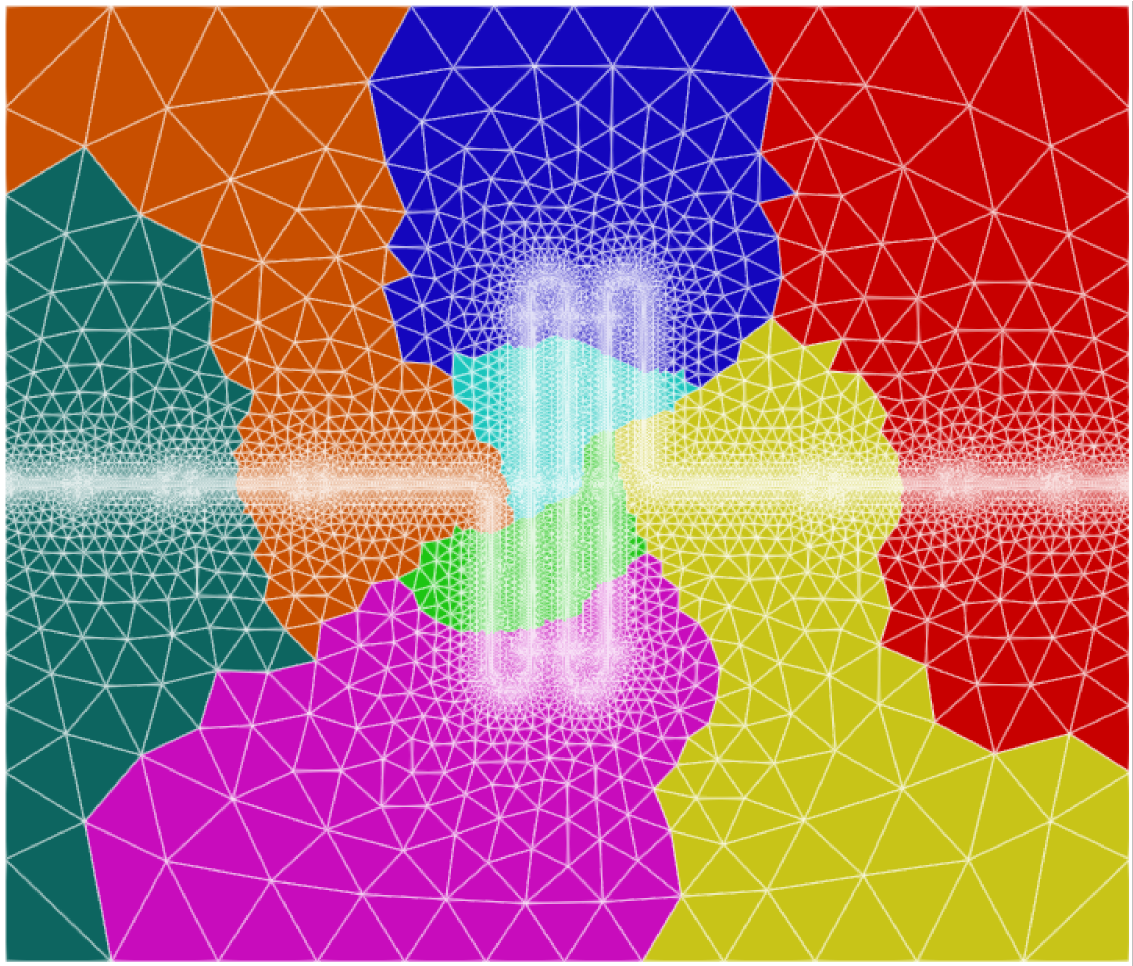}
    \caption{Example of a ``partitioned mesh'' of a coplanar waveguide resonator with airbridges distributed along the resonator. For visualization, only the mesh on the surface of the device is shown. The mesh is partitioned into eight sub-domains (denoted by different colors) with approximately the same number of tetrahedrons in each using METIS \cite{karypis1998fast}. Although the size of this mesh does not require a domain decomposition method to be utilized, such mesh partitioning is a critical step in such methods, as well as for other uses such as parallelizing preconditioners.}
    \label{fig:ddm-illustration}
\end{figure}

To more effectively utilize parallel computing architectures and alleviate some of the memory constraints, domain decomposition methods are commonly utilized \cite{jin2015finite}. Many domain decomposition strategies exist for solving PDEs with varying advantages and disadvantages, but the central idea is always to follow a ``divide and conquer'' approach that seeks to take the very large problem and break it up into smaller sub-problems that can be solved in parallel. The decomposition into these sub-problems is dictated by the mesh of the overall geometry, which is partitioned into different sub-domains as needed for the various algorithms. An illustration of this for a coplanar waveguide resonator with dimensions commonly employed in cQED devices is shown in Fig. \ref{fig:ddm-illustration}.

In modern domain decomposition methods, the way these sub-problems are ``tied'' together involves forming a global interface problem that must be solved. The size of this global interface problem varies depending on the method used, but generally its size is related to the number of basis functions that exist on the interface surfaces between the different sub-domains. The size of this interface problem can still be quite large for practical engineering analysis, so it will often need to be solved with an iterative method. Importantly, the dimension of the interface problem is much smaller than simply attempting to solve the entire large-scale problem directly, so the onset of the computer storage problems discussed above can be significantly delayed. 

Ultimately, there are important tradeoffs between the number of sub-domains, the number of basis functions inside each sub-domain, and the final size of the interface problem that must be solved; all of which must be considered in the context of the available parallel architecture being used for the simulation. Further, there are important considerations related to the conditioning of the interface problem and the overall parallel efficiency of the complete simulation that also play into how a domain decomposition method should be formulated or applied to a particular problem. Due to the complexities involved, we will not comment on these methods in depth, but will reiterate that they are a very powerful set of techniques for solving large-scale problems using distributed memory parallel computers. A more detailed introduction to these methods can be found in \cite{jin2015finite}.

Depending on the context the domain decomposition method is designed for, some simplifications can be utilized. For instance, one important application for domain decomposition is in the analysis of periodic structures like phased array antennas. In these systems, the same antenna element is repeated many times to form the overall array. If the feeding and beamforming network part of the array does not need to be included directly in the simulation, it is likely that all sub-domains in the simulation are identical. As a result, the mesh can be reused in each sub-domain so that the corresponding ``local'' matrix characterizing each sub-domain are identical. Then, the sub-domain matrix can be factorized once and reused everywhere, greatly reducing the memory requirements of the algorithm. Alternatively, if the sub-domain is too large, a single preconditioner can be formed and used for all sub-domain problems to still provide some speedup to the solution process. Other simplifications in the solution process of the global interface problem could also be sought. However, the requirement of completely identical sub-domains is often challenging to realize in many practical applications, limiting how broadly these methods can be utilized to achieve accurate analysis results. Despite this, there may be some use cases for these methods in analyzing grids of qubits in larger cQED chips.

In general, domain decomposition methods have reached a level of maturity where various forms of them can be found in commercial tools. However, the use of domain decomposition methods is primarily advantageous for exploiting the power of high-performance computing clusters that use a distributed memory architecture. Often, the licensing structures for commercial tools have steep pricing increases to leverage the full performance of such computing architectures. As a result, research groups may have limited opportunities to utilize these capabilities without making a significant investment with a particular commercial vendor (academic research groups may be able to benefit from reduced pricing, depending on the commercial vendor).

Before moving on, it is worth mentioning that domain decomposition methods are primarily formulated in the context of PDEs and so are relevant to FEM. However, some methods have also been developed for integral equations accelerated with fast algorithms (e.g., see \cite{li2007wave}), although these are not as extensively developed. Further, domain decomposition methods are primarily studied and applied to solving driven simulation problems. The direct application of these strategies to solving eigenvalue problems is not always possible; although, some limited research on domain decomposition methods for EM eigenvalue problems does exist (e.g., see \cite{liang2023two}). 

\subsection{Summary}
\label{subsec:cem-summary}
In current research practice, FEM is nearly the only CEM technique currently used for cQED device modeling. It is a natural question to ask whether this is the ``best practice'', or if other CEM techniques may have a role to play in this field. Throughout Section \ref{sec:cem_overview}, we have discussed the advantages and disadvantages of many different CEM techniques to help answer this question. Although a definitive answer applicable to all cQED devices is not possible, the challenges of modeling them do make FEM the most widely accessible and viable option for most researchers. However, well-developed MoM solvers accelerated with fast algorithms are also highly capable tools that should be explored more thoroughly for cQED modeling applications. Likewise, DGTD methods should be explored more thoroughly for modeling cQED devices. 

As will be discussed in more detail in Section \ref{subsec:higher-order-basis}, achieving accurate solutions for cQED devices often requires the use of higher-order basis functions that provide superior interpolation accuracy. As a result, even advanced FDTD methods using conformal meshes and implicit time-stepping algorithms to alleviate the stability constraint are still likely to struggle to provide accurate results in many cQED situations. However, CEM researchers who develop these methods could use cQED devices as a challenging application to help spur further developments in their techniques and potentially demonstrate their viability for these devices. These same challenges also likely limit the role that DEC solvers can play for modeling cQED devices. 

While SEMs can use higher-order basis functions, without progress on simple and user-friendly ways to automatically mesh practical cQED devices to leverage the capabilities of these methods they are unlikely to play a significant role for this application space. Whether higher-order basis functions are needed for integral equation solvers to achieve suitable accuracy in reasonable simulation times for these applications is an open question that warrants further investigation. This would also have direct implications on how likely PEEC methods are to achieve suitable performance, as these generally use simpler basis functions than typical MoM solvers. 

Overall, there are many opportunities for further research on CEM techniques to improve their performance for the challenges of modeling cQED devices. There are also many opportunities for research on cQED theory to devise alternative strategies for how the results of CEM simulations can be more effectively utilized in predicting quantum properties of interest to designers. More suggestions on these coupled research opportunities are discussed in Section \ref{sec:conclusion}.

\section{Eigenvalue Problems in Computational Electromagnetics}
\label{sec:eig-overview}
As mentioned in Section \ref{subsec:eigenvalue-problem-basics}, eigenvalue problems represent one of the most challenging problems to solve in numerical linear algebra. Correspondingly, all the general challenges with CEM methods discussed in Sections \ref{sec:cem_overview} and \ref{sec:practical-considerations} are also relevant to eigenvalue problems, and only serve to increase the difficulty already inherent in the problem. As a result, eigenvalue solvers have not been traditionally leveraged in EM applications as thoroughly as the more conventional driven simulations. Despite this, there are use cases for them, so they have been developed, albeit to a less mature degree than their driven counterparts.

In this section, we review some of the specific details of interest to eigenvalue problems in CEM. We first discuss in Section \ref{subsec:pde-eig} the standard approach of using FEM to discretize and solve these problems. Following this, we discuss in Section \ref{subsec:ie-eig} an alternative integral equation formulation that can also be utilized for these problems. We conclude in Section \ref{subsec:td-eig} by discussing how time domain methods can also be used to find eigenpairs.

\subsection{Partial Differential Equation Formulation}
\label{subsec:pde-eig}
The most common approach to formulating an eigenvalue problem in CEM is from the wave equation given in (\ref{eq:vec5}), but without a source function. In this case, the wave equation can be rearranged into the form of an eigenvalue problem as
\begin{align}
    \nabla \times \mu_r^{-1} \nabla\times\mathbf{E}_n  =  k_{0,n}^2 \epsilon_r  \mathbf{E}_n, 
\end{align}
where $k_{0,n}^2$ serves the role of the eigenvalue and $\mathbf{E}_n$ is the eigenvector. This equation can be discretized using FEM to arrive at a generalized eigenvalue problem given by
\begin{align}
    [S]\{v_n\} = \lambda^2_n [M] \{v_n\},
    \label{eq:pde-gen-eig1}
\end{align}
where the definitions of $[S]$ and $[M]$ were given in (\ref{eq:stiffness}) and (\ref{eq:mass}). 

One of the challenges in solving this system (beyond the standard challenge of large eigenvalue problems) is that the matrix $[S]$ that discretizes the curl-curl operator of the wave equation supports a large and non-trivial null space, which correspondingly causes it to also have an exceptionally large condition number. This null space does not preclude the use of shift-invert techniques to target certain eigenvalues in the spectrum, but it does still complicate the solution of the problem and can degrade performance \cite{hiptmair2002multilevel}. 

As a result, research has been done on specialized treatments to counter the ill effects of this null space, albeit not at a significant level in comparison to other topics within CEM. Broadly, the main approaches look at modifying the standard iterative eigenvalue problem solvers used in EM applications, typically the Lanczos or implicitly-restarted Arnolodi methods \cite[Ch. 4]{lehoucq1998arpack}, to constrain the solution space in some way. For example, in \cite{lee1991full} the Lanczos algorithm is modified to maintain orthogonality to the subspace of eigensolutions associated with a null eigenvalue for the application of finding propagating modes in a waveguide. Alternatively, some kind of preconditioning strategy can also be introduced to try and circumvent these problems. Despite these efforts, the solution of large EM eigenvalue problems still is an inherently challenging problem that could be more sensitive to errors than driven simulations.

One important point to note regarding these generalized eigenvalue problems is the boundary conditions for which an eigenvalue problem can typically be formulated. Generally, one assumes that the outer surface of the system being modeled is either a PEC, a perfect magnetic conductor (PMC), or a periodic boundary condition (PBC). Under these conditions, the system is ``closed'' and the PDE encoded in the wave equation can be shown to be a Hermitian operator that correspondingly supports a complete set of orthogonal eigenvectors with real-valued eigenvalues. The hermiticity of the problem also translates into the FEM-discretized matrix equation, which enables the use of eigenvalue problem solvers that can exploit this prior knowledge about the structure of the matrix. 

Unfortunately, if loss is incorporated into the system (e.g., via dielectric loss, lumped resistors, or absorbing boundary conditions), the hermiticity is lost and the solution of the eigenvalue problem becomes more complex. In particular, the additional $[Z]$ matrix in (\ref{eq:fem-fd-full}) that depends on $k_{0,n}$ rather than $k_{0,n}^2$ is introduced, causing the problem to become the quadratic eigenvalue problem 
\begin{align}
    [S]\{v_n\} + \lambda_n [Z] \{v_n\} - \lambda_n^2 [M] \{v_n\} = 0.
\end{align}
This quadratic eigenvalue problem can attempt to be solved with specialized eigensolvers \cite{bai2005soar}, or the quadratic eigenvalue problem can be reduced back to a linear generalized eigenvalue problem at the cost of introducing auxiliary unknowns that (at least) doubles the dimension of the matrix system being solved \cite{venkatarayalu2011efficient,zekios2015eigenanalysis}. Although this linearization does not significantly increase the cost of multiplications needed to be performed in the basic matrix-vector products of the solution procedure, the increased length of the eigenvectors that must be stored and manipulated will increase storage costs and accelerate when implicit restarting is necessary, making the problem more difficult to solve. 

In some instances, if the effect of the loss is expected to be small, a perturbation approach can instead be pursued to account for the loss \cite{zekios2014finite}. However, this is not a general-purpose approach that can be relied on in all scenarios. Regardless, because the problem is no longer Hermitian, the eigenvalues become complex valued and the orthogonality of the eigenvectors is lost. Ultimately, the presence of loss takes an already highly challenging numerical linear algebra problem and only makes it more difficult to solve reliably.

\subsection{Integral Equation Formulation}
\label{subsec:ie-eig}
In CEM, eigenvalue problems are almost always solved using FEM. However, some research has been performed on integral equation formulations as an alternative (e.g., see \cite{arcioni1995new,gil2006analysis,mira2005fast,navarro2022wide}). Although different names for these approaches have been used over the years, they are now predominantly referred to as the boundary integral-resonant mode expansion (BI-RME) method. The main interest in using such methods is to enable the use of surface discretizations (and eventually fast algorithms) to handle larger regions than can be readily accommodated with FEM. Unfortunately, the formulation and implementation of these integral equations is more involved than FEM, and so their development and testing has been much more limited. As a result, their complete properties for challenging EM modeling applications and for large geometries where fast algorithms would be needed have not been thoroughly shown.

The basic idea of the BI-RME method is to utilize a Green's function that is analytically known for a problem ``similar'' to the one intended to be solved rather than using the standard ``free-space Green's function'' typically used in integral equation derivations. For instance, an early application of the method considered enclosing the geometry to be studied inside a spherical cavity so that the Green's function of the empty spherical cavity could be used \cite{arcioni1995new}. Although this may seem counter-intuitive at first, the integral equation that is formulated can still find the eigenmodes of the ``internal'' geometry that was intended to be modeled. The main benefit of the BI-RME strategy is that the relevant Green's functions can then be broken into a quasi-static piece and a ``high-frequency correction'' written in terms of resonant modes of the system the Green's function is valid for (i.e., for the case mentioned earlier, the resonant modes of the spherical cavity). Decomposing the Green's function in this way enables the derivation of a linear eigenvalue problem, whereas a direct use of the free-space Green's function would lead to an implicit eigenvalue problem that is highly cumbersome to solve. 

One benefit of the BI-RME method is that the matrix system that is formed only requires discretizing the part of the system geometry where the analytical Green's function does not already satisfy the boundary conditions. If a good analytical Green's function can be found that is close to the geometry to be modeled, this can allow for a significantly smaller matrix system to be formed than would be needed for a FEM discretization of the same modeling problem. For instance, if a system to be modeled is to be naturally enclosed in a rectangular cavity, the analytical Green's function within a rectangular cavity could be used to greatly lower the dimension of the matrix system \cite{mira2005fast}. 

Although this can be a benefit, it comes at the cost of a significantly labor intensive process due to the complexity of these formulations for different Green's functions. For example, the mathematics must be redone, the new expressions may present difficulties for efficient and accurate numerical integrations, and the code must be substantially rewritten. Further, it may be possible that an analytical Green's function will not decompose as advantageously into quasi-static and high-frequency corrections as desired to arrive at a linear eigenvalue problem. Hence, the implementation of this BI-RME method for a particular application has some interesting and appealing possibilities, but there will likely be non-trivial challenges that must be solved along the way.

\subsection{Time Domain Eigenpair Extraction Approach}
\label{subsec:td-eig}
Another option for finding eigenpairs is to use time domain methods (e.g., see \cite{banova2013eigenvalue,banova2013performance,dawson2018application}). In this approach, a time domain simulation is excited with an input waveform covering the frequency bands of interest and the fields are stored at some number of points in the computational domain. These recorded time domain signals can then be post-processed using various signal processing techniques to identify the eigenvalues associated with the observed resonant frequencies. In its simplest form, this can involve taking a Fourier transform and identifying peaks in the spectrum. Alternatively, more sophisticated signal processing methods can be utilized to obtain better accuracy and enhance the ability of the method to resolve closely spaced eigenvalues. To enhance robustness, it is generally necessary to sample and store the fields at multiple points in the simulation domain to avoid missing any eigenvalues that correspond to a resonant mode that happens to have a null (or simply a much smaller value) at the particular sampling point selected.

In the basic procedure described above, only the eigenvalues are found. To also compute the eigenmodes in a memory-efficient way, a second time domain simulation can be performed once the eigenvalues of interest have been found \cite{dawson2018application}. As the solution is advanced in time, ``running DFTs'' can be computed for all the basis functions used in the discretization for each eigenvalue of interest. The end result is the complex-valued eigenmodes for each eigenvalue of interest.  The memory storage requirements of this approach is equivalent to storing another vector for each eigenmode being computed, which in general should be manageable and (sometimes substantially) less than would be required by a traditional iterative eigensolver algorithm \cite{banova2013performance}.

Hence, when a very large number of eigenpairs are desired to be found, a time domain approach has some compelling advantages. Another benefit of this approach is that the underlying time domain solver is the same as that for a standard driven simulation, making any advances to the solver done to enable better eigenpair extraction also able to improve performance for doing standard driven simulations (e.g., extracting scattering or impedance parameters). Likewise, advanced methods like domain decomposition and other parallelization efforts for the underlying time domain solver can be directly leveraged for eigenpair extraction. This is in contrast to the more traditional eigensolver approach as described in Section \ref{subsec:pde-eig}, where such methods are not necessarily directly compatible with the solution of the generalized eigenvalue problem. 

Although these are interesting properties, it is important to note that these time domain approaches have not been as thoroughly tested for these applications and are not without their challenge for performing long simulations (which can be needed to achieve a specified frequency resolution). Strategies exist to accelerate this kind of process as well (e.g., see \cite{nayak2024fly} and the references therein), but their underlying accuracy for challenging simulation conditions are also not as thoroughly tested. 

\section{Practical Considerations for cQED Device Modeling}
\label{sec:practical-considerations}
With a broad overview of CEM methods now covered, we turn our attention to discussing some more practical topics that are of particular importance when trying to get a numerical method to work ``well'' for a specific modeling problem. We will illustrate these concepts in the broad context of modeling typical cQED devices; although, many of the considerations are also relevant for other EM modeling applications that have complex features that can stress the performance of typical CEM tools. More specifically, we will discuss the use of higher-order basis functions, adaptive mesh refinement, and challenges with modeling multiscale structures in this section. Our discussion will be primarily focused on FEM implementations due to its prevalence in current cQED research, but many of the general considerations also apply to MoM simulations.

\subsection{Higher-Order Basis Functions}
\label{subsec:higher-order-basis}
\begin{figure}[t!]
	\centering
	\includegraphics[width=0.93\linewidth]{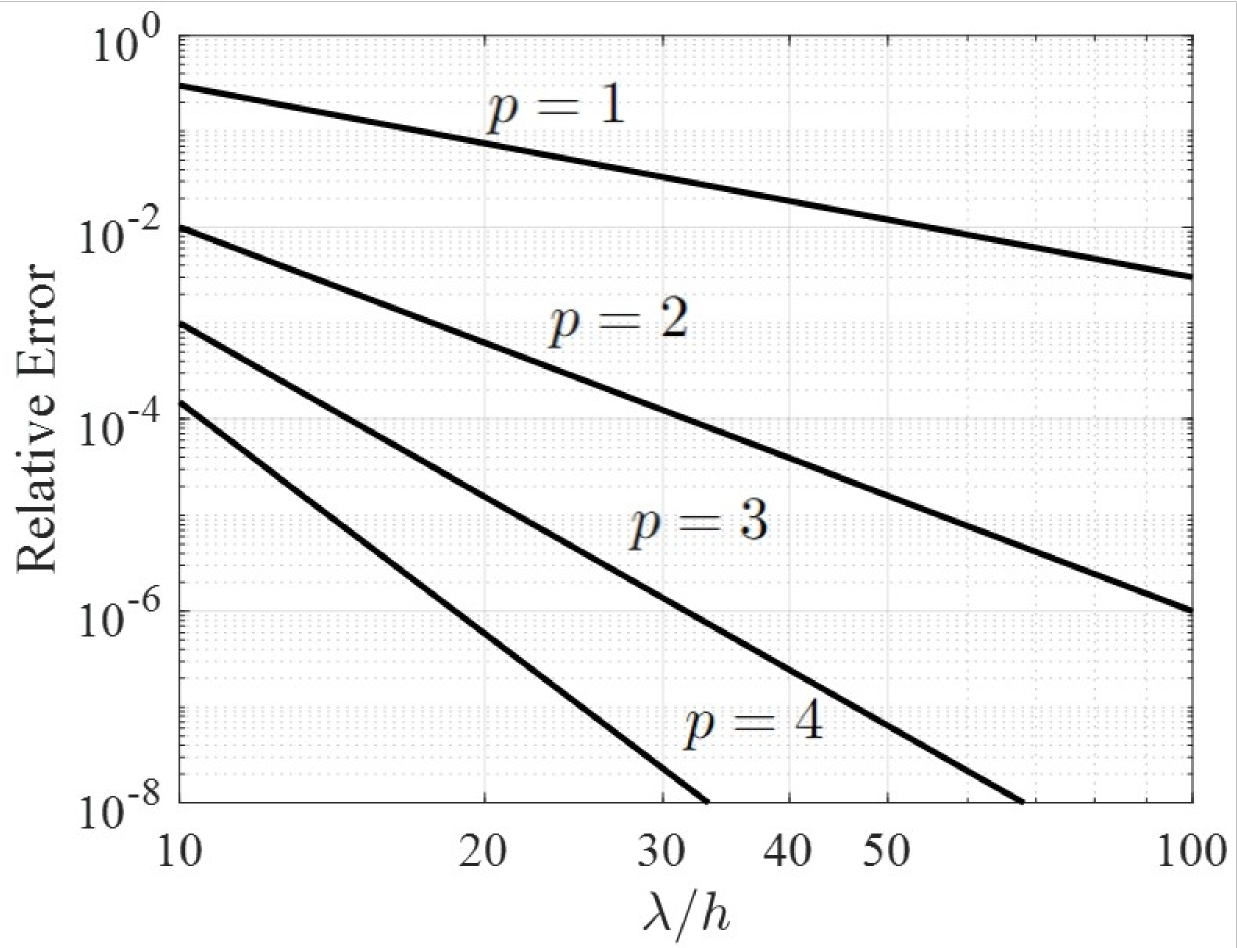}
	\caption{Comparison of notional convergence rates for a 1D FEM code using different orders of polynomial interpolation (denoted by $p$) in their basis functions as a function of average element length $h$ (adapted from \cite{scott2002errors}).}
	\label{fig:1d-phase-error}
\end{figure}

The most basic form of vector basis functions used in FEM support a first-order-like representation of the fields to be solved for (i.e., approximately linear variations across a mesh element). However, higher-order versions of this kind of basis function have been developed, and are crucial for achieving high accuracy in reasonable computation times for many practical applications. For simplicity, we show in Fig. \ref{fig:1d-phase-error} the performance comparison in 1D for various FEM discretization orders, but the basic trend still holds in 3D meshes. From Fig. \ref{fig:1d-phase-error}, it is clearly seen that increasing the polynomial order of basis functions can significantly increase the convergence rate compared to only making the mesh elements smaller. An analysis of the numerical dispersion error provides the important takeaway that the convergence rate versus basis function order scales as $(h/\lambda)^{2p}$, where $h$ is the average mesh element length, $p$ is the polynomial order, and $\lambda$ is the wavelength \cite{jin2015finite,scott2002errors}. 

As a result, there are generally three strategies that are employed to improve the accuracy of an FEM solution. The first is known as $h$-refinement, which involves making the mesh elements smaller so they can more accurately interpolate the EM fields. The second is known as $p$-refinement, which involves increasing the polynomial order of the interpolating basis functions. Finally, there is $hp$-refinement, which involves a combination of $h$- and $p$-refinement strategies. Generally, performing this kind of refinement ``by hand'' is highly impractical, so CEM tools focus on performing adaptive mesh refinement that requires as little interaction from the user as possible. These concepts will be discussed in more detail in Section \ref{subsec:amr}.

\begin{figure}[t!]
	\centering
	\includegraphics[width=0.95\linewidth]{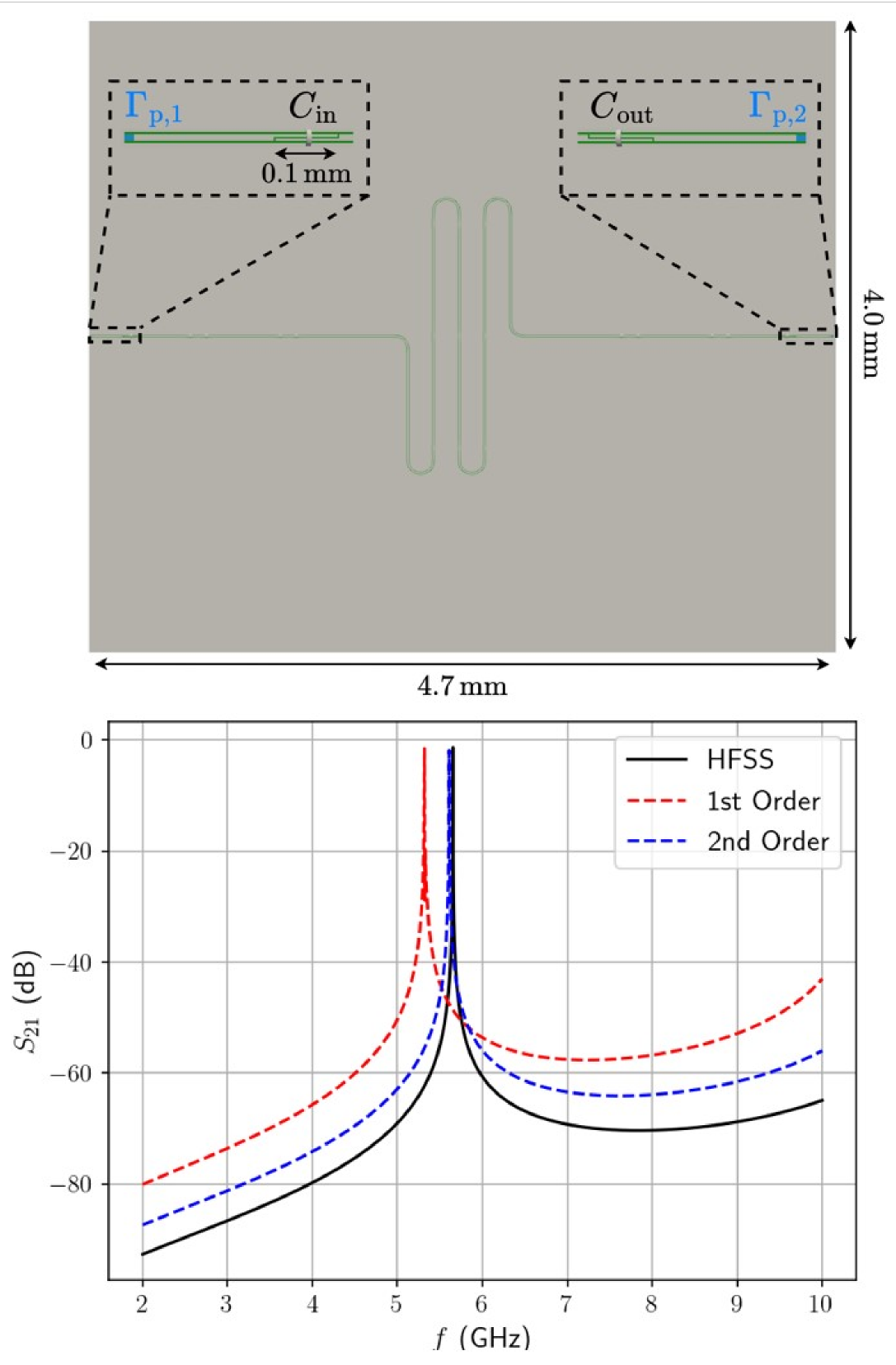}
	\caption{(Top) Schematic of a coplanar waveguide resonator with dimensions of typical cQED devices. The substrate has a relative permittivity of $\epsilon_r = 11.0$ and is $0.5\, \mathrm{mm}$ thick. (Bottom) Frequency response computed using first- or second-order basis functions with the mesh illustrated in Fig. \ref{fig:ddm-illustration}. As a reference, HFSS simulations were also performed with adaptive $h$-refinement using third-order basis functions.}
	\label{fig:cpw-plus-freq-response}
\end{figure}

As a more concrete example, we show in Fig. \ref{fig:cpw-plus-freq-response} how changing between first- and second-order basis functions throughout the entire mesh (i.e., an example of uniform $p$-refinement) can significantly improve the accuracy of the frequency response of a typical cQED resonator. The mesh used for this example is shown in Fig. \ref{fig:ddm-illustration}, although no partitioning of the mesh or domain decomposition was actually used for these simulations. We also show the results of a commercial simulation tool (Ansys HFSS) that uses a different mesh and third-order basis functions in Fig. \ref{fig:cpw-plus-freq-response} as a reference solution. This mesh was generated following an adaptive $h$-refinement process, enabling a more robust solution in most cases than pure $p$-refinement. As a result, most commercial or open-source tools provide the users with the option to perform adaptive $h$-refinement or $hp$-refinement. 

Before continuing, it should be mentioned that there are different conventions for referring to the order of basis functions in CEM. These conventions arose for various reasons, but a central aspect is that for a single vector basis function the variation of the fields in the direction of the vector is one order less than the variation in the direction normal to the vector. For example, for the vector basis functions illustrated in Fig. \ref{fig:edge-elements}, the component tangential to the edge with which the basis function is associated has a constant value (zeroth-order accuracy), while the normal component has a linear variation (first-order accuracy). Further, analysis of FEM for CEM applications has shown that the overall solution accuracy strongly depends on the accuracy of the curl of the vector field, not just the interpolation order for the vector field itself \cite{jin2015finite}. Hence, if we use first-order polynomials in our vector basis function, the curl will only have zeroth-order accuracy. Thus, some authors denote the order of the basis function based on what the lowest accuracy can be guaranteed to be while other authors choose to denote the order based on the highest accuracy that can be achieved (for certain components of the field). Throughout this document, we follow this second convention so that the lowest-order basis function that can be used for FEM is a first-order one. Notably, Ansys HFSS follows the opposite convention, so what we refer to here as a third-order function would be denoted as a second-order one in that tool.  

Although higher-order basis functions are typically very beneficial, there are some drawbacks. For instance, near a field singularity or other regions where the fields change in less of a smooth manner a higher polynomial order can actually do a poor job in representing the underlying field variation \cite{jin2011theory}. In situations like this, it is generally better to use smaller elements with lower-order basis functions or to utilize specialized basis functions meant to interpolate singular fields \cite{peterson2016basis}. Hence, the idealized convergence properties of $(h/\lambda)^{2p}$ may only be fully seen in practice on simple geometries. Additionally, higher-order basis functions also tend to progressively worsen the condition number of the matrix, which can complicate the iterative solution of large systems unless suitable preconditioners can be devised. This is especially true for the higher-order basis functions that can be used in $p$-refined or $hp$-refined FEM \cite{jin2015finite}. 

It is also worth mentioning that the total number of basis functions per mesh element also rapidly grows with the increasing polynomial order. For instance, on a tetrahedron, the total number of basis functions scales as $O(p^3)$, where $p$ is the polynomial order \cite{jin2015finite}. More specifically, a first-order discretization would have 6 basis functions per element, a second-order discretization would have 20 basis functions per element, and so on, with a fifth-order discretization having 140 basis functions per element. Ultimately, the higher-order discretizations can lead to a rapid growth in the size of the matrix while also reducing its sparsity. Due to these challenges, many general-purpose FEM codes that are meant to be as robust as possible do not attempt to use very high-order basis functions. For example, they may stop at roughly third order to focus on the robustness of the tool for general use cases. 

\begin{figure}[t!]
	\centering
	\includegraphics[width=0.7\linewidth]{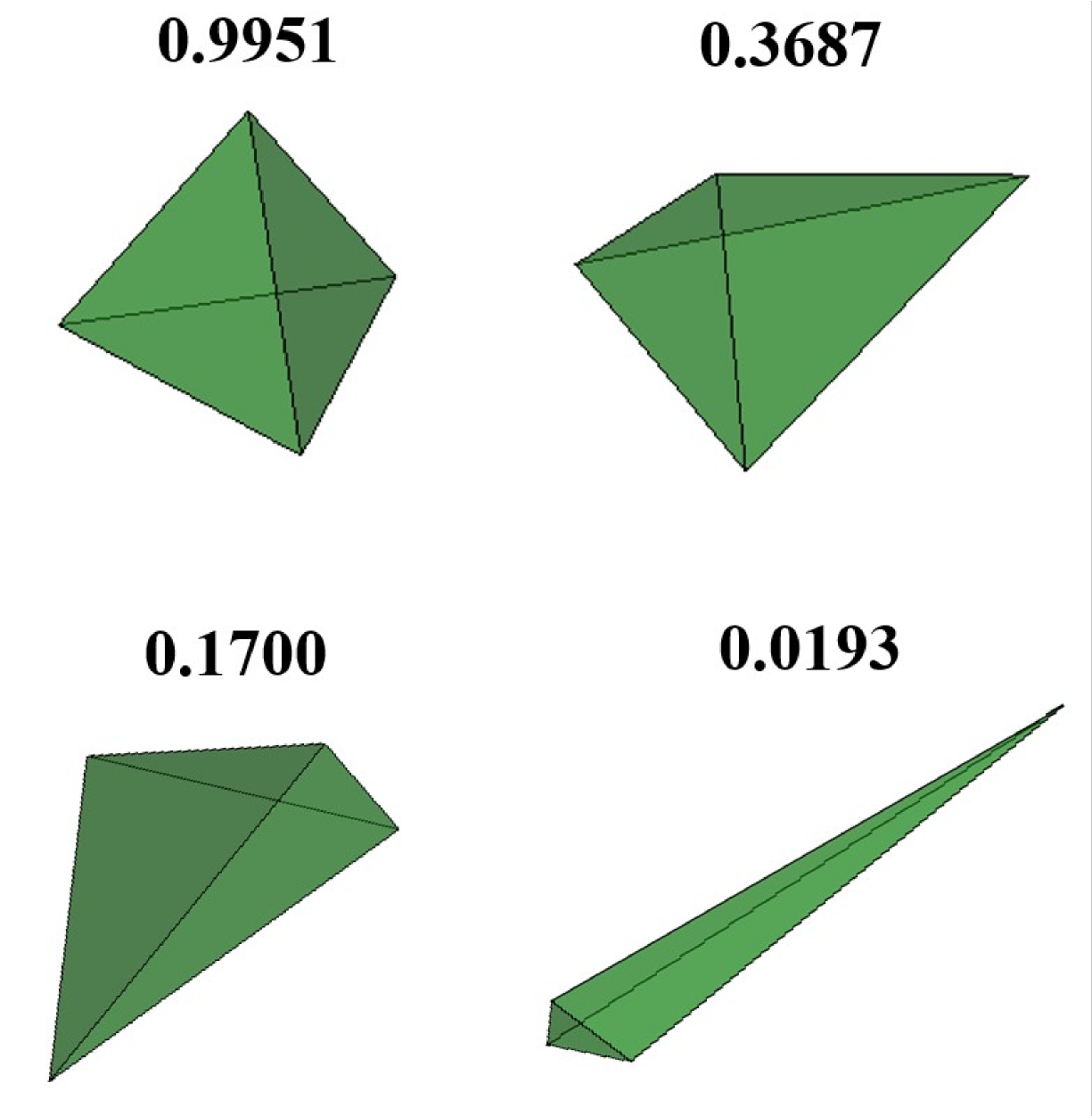}
	\caption{Examples of tetrahedrons with different levels of quality. The quality numbers shown above each tetrahedron are computed using the ``scaled Jacobian'' metric from the Coreform Cubit meshing software, for which a target ``acceptable range'' of $0.2$ to $1$ is suggested. The full range of possible values for this metric is $-1$ to $1$.}
	\label{fig:mesh-quality}
\end{figure}

Many of these considerations can be further exacerbated for meshes that contain low \textit{quality} elements \cite{shewchuk2002good}. Although various ways exist to quantify the quality of a mesh element, generally a ``good'' quality element will have a fairly uniform aspect ratio for the different sides and angles of the element. A comparison of good and bad quality elements is shown in Fig. \ref{fig:mesh-quality}. Generally, we want good quality elements because this leads to the interpolation functions used in our basis functions performing better over the range of the element and also improves the accuracy of derivatives computed from the basis functions \cite{shewchuk2002good}. The impact of mesh quality on these considerations is typically amplified for higher-order basis functions, and lower-quality elements also increase the ill-conditioning associated with higher-order basis functions \cite{jin2015finite}. Since realistic geometries can often make it very difficult to always achieve high-quality mesh elements, this also contributes to why many FEM codes do not use extremely high-order basis functions. 

In addition to higher-order basis functions, higher-order meshes can also be used, which are typically referred to as curvilinear elements. As an example, a standard tetrahedral mesh would have straight linear edges. A curvilinear tetrahedral element would still have six edges, but the edges could follow a quadratic or a higher-order path. These elements can be helpful in more accurately representing the geometrical features of curved surfaces using fewer elements. However, this comes at the cost of more complicated mesh generation, user pre-processing, and implementation of the FEM code. Further, one can run into some similar problems as those discussed with higher-order basis functions that complicates the robustness of these methods for practical problems that can become quite complex. Overall, these features are available in some commercial tools, but do not seem to be significantly utilized in most applications.

\begin{figure}[t!]
	\centering
	\includegraphics[width=0.9\linewidth]{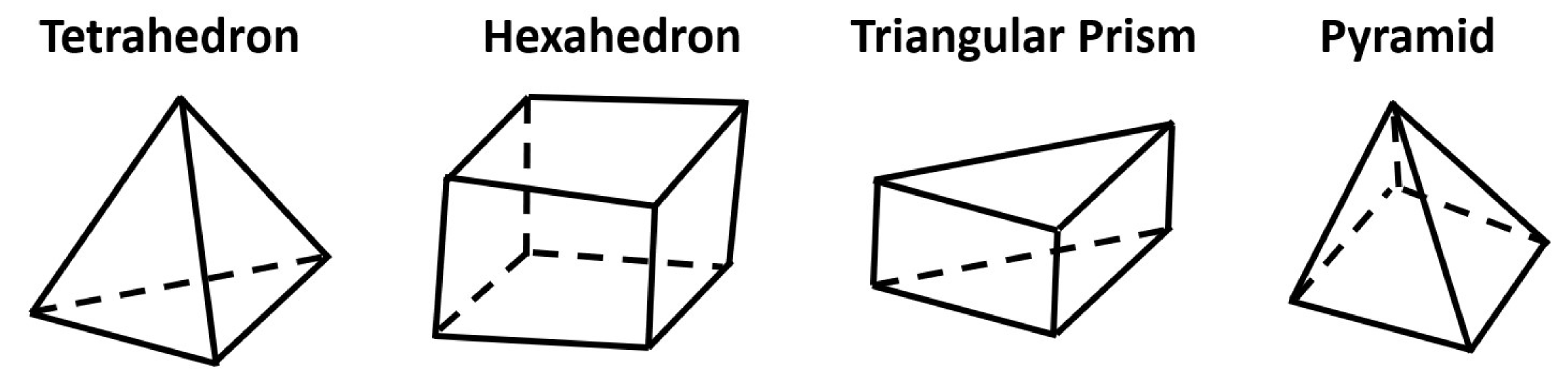}
	\caption{Different kinds of 3D mesh elements that are commonly used in computational analysis.}
	\label{fig:mesh-cells}
\end{figure}

From these general discussions, it is clear that mesh quality plays an important, albeit difficult to quantify, role in affecting the performance of an FEM tool. To help achieve more high-quality mesh elements, certain applications can benefit from exploring the use of less frequently employed mesh elements. For instance, most 3D FEM codes only use tetrahedral meshes, but other options exist, such as hexahedral and triangular prism meshes, as shown in Fig. \ref{fig:mesh-cells} \cite{jin2015finite}. 

Geometries with thin layers can potentially benefit from using a triangular prism mesh in certain regions, as these elements help decouple the vertical and longitudinal variations of the mesh and corresponding basis functions \cite{jiao2007layered,jin2015finite}. This could be particularly useful for 3D analysis of cQED devices that explicitly model the thickness of the superconductors and other interface layers, which may be important for improving the reliability and accuracy of surface participation ratio calculations. However, practical simulations generally require these triangular prisms to only be used in part of the mesh, requiring more complicated pre-processing from the user and FEM code implementation to handle working with multiple mesh types. Further, the extremely thin layers present in cQED devices will exacerbate multiscale modeling challenges, which will be discussed more in Section \ref{subsec:multiscale}.

\subsection{Adaptive Mesh Refinement}
\label{subsec:amr}
As alluded to earlier, there are generally three strategies to improve the accuracy of an FEM analysis: $h$-refinement, $p$-refinement, or $hp$-refinement. In practice, most implementations will focus on $h$-refinement with some fixed polynomial order used throughout the entire mesh or alternatively follow a complete $hp$-refinement strategy. In each case, adaptive mesh refinement algorithms are typically favored that are meant to balance the needed simulation time to achieve a desired level of solution accuracy. These adaptive algorithms begin with a coarse mesh, solve the problem, predict the error across the mesh, and then use this error prediction to determine which elements should be refined. After performing the refinement, the problem is solved again, and the process is repeated until some specified convergence criteria are met. 

\begin{figure}[t!]
    \centering
    \includegraphics[width=0.8\linewidth]{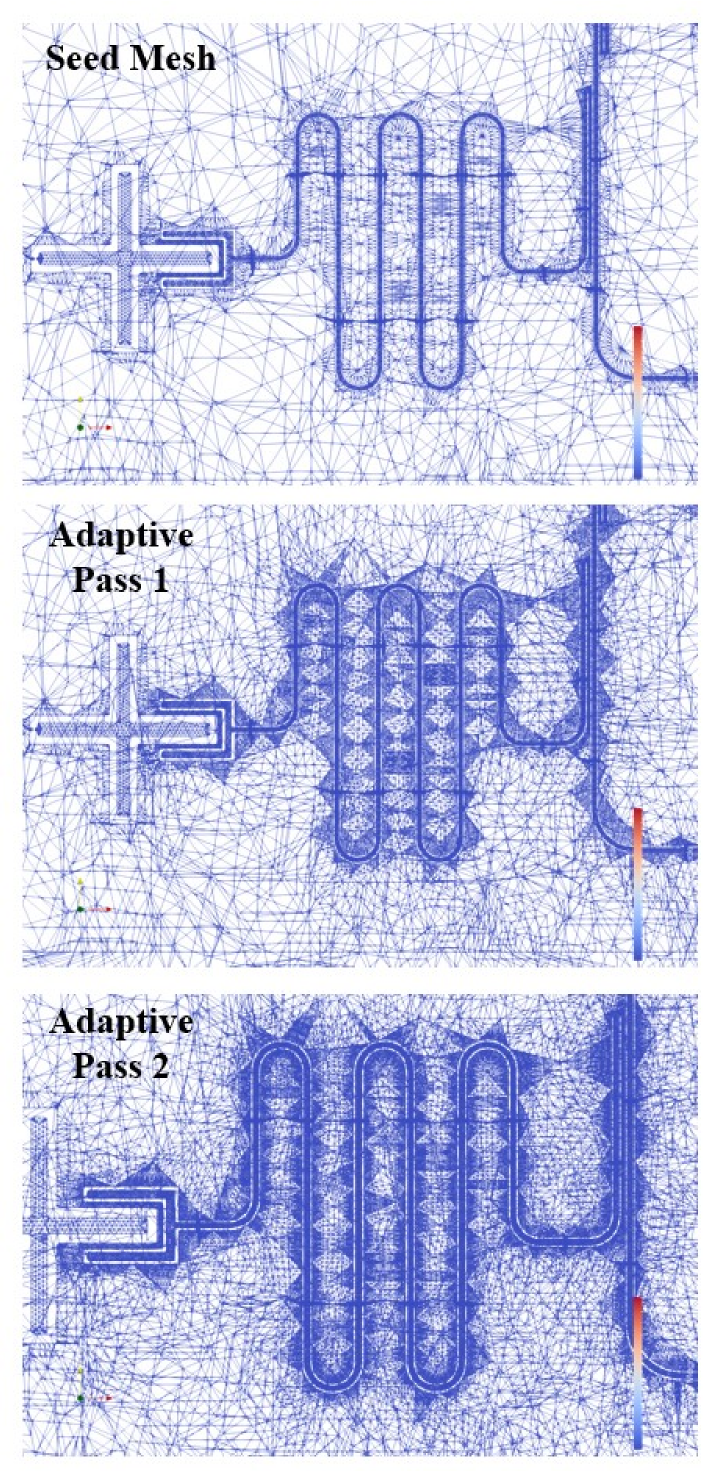}
    \caption{Example of $h$-refinement for a cQED structure in a tool that lacks integration between the geometry creation and mesh refinement software (specifically, AWS Palace v0.13.0). The low-quality elements in the seed mesh remain low-quality elements as the mesh is refined, limiting the utility of this approach.}
    \label{fig:mesh-refinement}
\end{figure}

An example of an $h$-refinement process is shown in Fig. \ref{fig:mesh-refinement} for a cQED qubit and resonator. This refinement was performed with a tool that did not have integration between the geometry creation and mesh refinement software (specifically, AWS Palace v0.13.0). As a result, the only way to refine the mesh elements is through a simple bisection process that splits tetrahedrons into smaller ones. This is not ideal, because if the beginning ``seed'' mesh has low-quality elements the bisection refinement process will improve the mesh quality very slowly. Generally, commercial FEM tools will have tight integration between the geometry creation and meshing software so that at each refinement pass the mesh can be remade in a more flexible manner. This helps the refined mesh better resolve the geometric features and generally also reduces the number of low-quality mesh elements that are generated. 

Another key component of the mesh refinement process is the error predictor, with many different approaches being available and having their own advantages and disadvantages \cite{sun2000adaptive,becker2001optimal,nicaise2005zienkiewicz,harmon2020adjoint,wang2023posterior}. In their simplest forms, error predictions can be made by monitoring the residual of some quantity that is known to physically have certain properties that will not be exactly preserved by a FEM discretization. For instance, one can compute the magnetic field by taking the curl of the electric field that was actually solved for. Physically, we know the tangential magnetic field should be continuous between mesh elements, but this will not be exactly preserved in the discrete setting using typical basis functions. The degree to which this property is violated can then be used as part of a predictor of the local error in a mesh element \cite{sun2000adaptive}. 

Other more involved procedures also exist, such as those that involve solving the adjoint problem to predict errors \cite{becker2001optimal,harmon2020adjoint,wang2023posterior}. Broadly, adjoint-based methods seek to leverage the mathematical connection between the solution of standard ``forward'' FEM problems and the solution of their ``dual'' problems (i.e., the ``adjoint'' problem) to efficiently provide information about the sensitivity of particular model outputs to model inputs. 

A key property of adjoint-based methods is that they can be goal-oriented; i.e., they can specifically focus their efforts towards specified quantities of interest (QoIs). For instance, in the case of adaptive mesh refinement, an adjoint method will allow the automated assessment of how the predicted error across the mesh will influence the specified QoIs. This allows the mesh to then be refined in the locations of the model that will most rapidly improve the accuracy of the solution for the QoIs. This is in contrast to conventional error predictors that are not directly tied to a QoI, which can typically lead to unnecessary ``over-meshing'' in many areas of the model and a potentially less reliable adaptive process overall. Exploring the use of these more advanced adaptive mesh refinement algorithms for particularly challenging QoIs in cQED modeling could provide significant improvements to the robustness of FEM analysis for these applications.

As a final note on adaptive mesh refinement, if one is using $hp$-refinement, different criteria have to be used to decide \textit{how} an element should be refined. For example, a particular element could have just $h$- or $p$-refinement in a pass of the algorithm, or both could simultaneously be performed on an element. Due to the complexity of some of the heuristics involved in this process, even though $hp$-refinement would conventionally be considered the best simulation approach, it can be possible for the $hp$-refinement algorithms to not be as robust as a simpler approach like $h$-refinement for challenging applications. We have seen this occur in our own experience in modeling certain cQED structures; however, this is not always the case, so we do not uniformly recommend users avoid utilizing $hp$-refinement methods in this application space. Rather, there could be an opportunity for further research on these methods to improve their reliability for a wider range of applications than has perhaps been focused on in the past.

\subsection{Multiscale Modeling Challenges}
\label{subsec:multiscale}
As discussed previously, multiscale geometries for EM modeling purposes can generally be thought of as simultaneously having wavelength to highly sub-wavelength sized features. To accurately resolve these features while meshing the geometry, element sizes can need to vary by several orders of magnitude throughout the geometry. When used in FEM, such a multiscale mesh typically produces a highly ill-conditioned matrix. To understand why this occurs, it first helps to discuss the origin of the ``low-frequency breakdown'' of (\ref{eq:fem-fd-full}). As the frequency is reduced, the $[M]$ term becomes less and less dominant to the $[S]$ term, until eventually the problem becomes unsolvable due to the large null space of $[S]$ \cite{zhu2010theoretically}. Ultimately, what matters is the size of the mesh elements relative to the wavelengths involved, which is why this is also sometimes referred to as the ``long-wavelength breakdown'' effect. Strategies to address this breakdown problem are an area of substantial interest, and the most common of such methods is tree-cotree splitting (TCS) \cite{wang2010application, lee2003hierarchical}, which is available in some commercial FEM tools. 

Unfortunately, while TCS vastly outperforms traditional methods when $[S]$ dominates over $[M]$, it suffers from relatively poor performance at ``high frequencies'' when $[M]$ dominates over $[S]$ \cite{elkin2025improvements}. As a result, it can be challenging to determine when using TCS will be beneficial. This issue is exacerbated for multiscale meshes, as these can be viewed as simultaneously having areas of the mesh where $[S]$ dominates over $[M]$ and others where $[M]$ dominates over $[S]$. Ultimately, this leads to an ill-conditioning of the overall matrix equation that can be challenging to deal with whether using TCS or traditional discretization methods. Recently, new strategies for overcoming these challenges have begun to be investigated that formulate wave equations in terms of EM potentials rather than fields \cite{li2016finite,yan2021continuous,elkin2025improvements}. 

\begin{figure}[t!]
    \centering
    \includegraphics[width=0.9\linewidth]{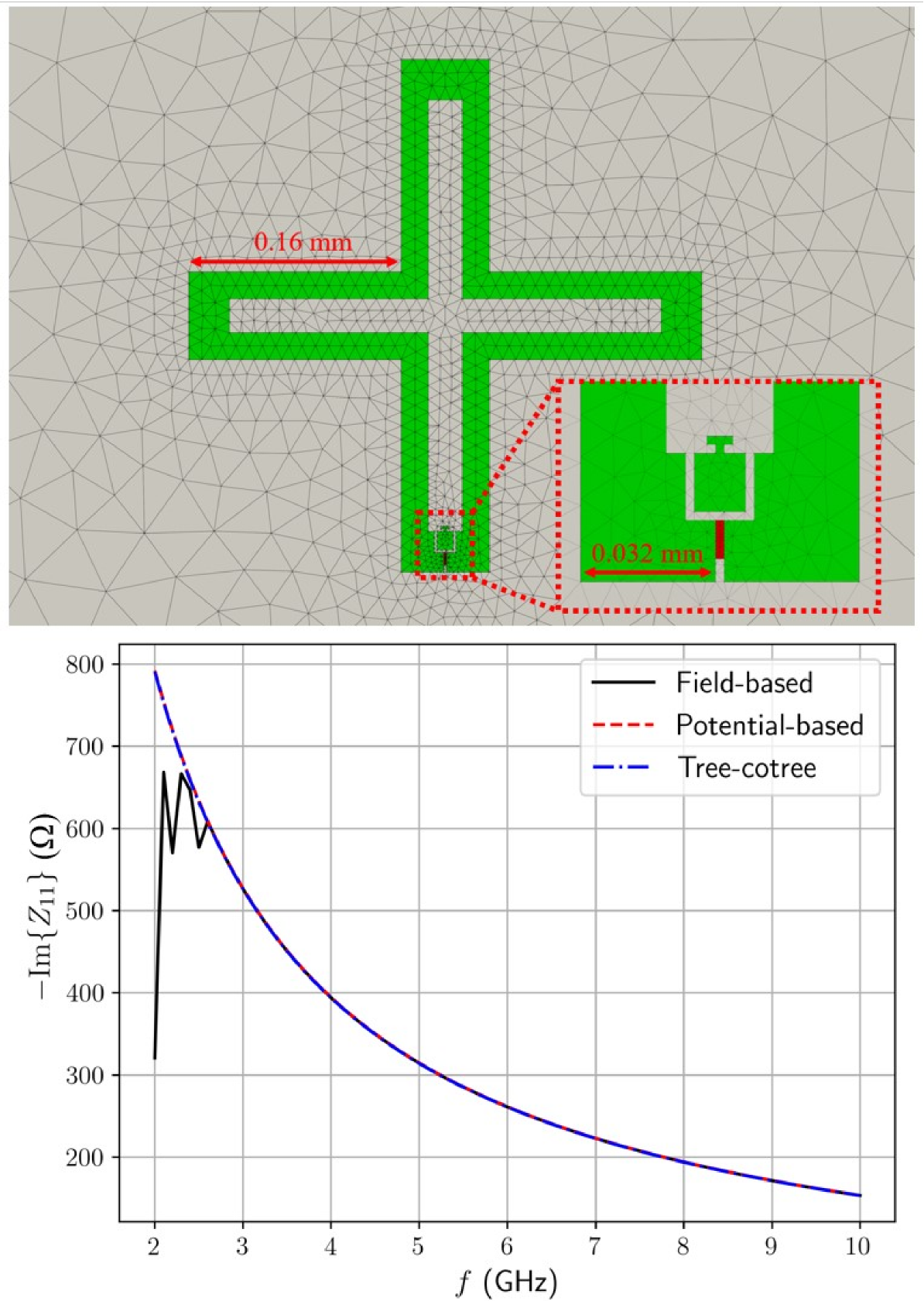}
    \caption{(Top) Part of the mesh of an isolated transmon qubit where the $175 \, \mathrm{nm}$ thick layer of the thin-film is modeled explicitly. The inset shows the features close to where the Josephson junctions would be located in the qubit. For simplicity, the simulation is driven with a single lumped port shown in red rather than including multiple ports to represent the different junctions in the SQUID loop. The full substrate of the model is $1.4 \times 1.1 \times 0.5 \, \mathrm{mm}^3$ and has a relative permittivity of $\epsilon_r = 10$. The bounding box for the simulation has a PEC boundary condition and is flush with the outer extents of the substrate in the plane of the qubit and has an additional $0.1 \, \mathrm{mm}$ ($0.2 \, \mathrm{mm}$) region of air above (below) the qubit chip. (Bottom) Imaginary part of the input impedance of the qubit computed using a traditional field-based FEM formulation, a potential-based method that is more robust for multiscale structures, and TCS. The field-based method breaks down completely below approximately $2.6 \, \mathrm{GHz}$ while the other methods remain stable at all frequencies.}
    \label{fig:xmon-thin}
\end{figure}

To highlight these considerations in more detail, here we look at the solver performance for geometries of varying simulation difficulty in the frequency and time domains using traditional field-based FEM as presented in Section \ref{subsec:fem}, TCS, and the potential-based method of \cite{elkin2025improvements}. To begin, we consider in the frequency domain how the thin layers common in cQED devices can cause issues for conventional simulation approaches. More specifically, we consider an isolated transmon qubit shown in Fig. \ref{fig:xmon-thin} where we explicitly model the $175 \, \mathrm{nm}$ thickness of the superconducting thin film; however, to keep the focus of the simulation effects purely on the thinness of this layer, we treat the superconductor as an impenetrable PEC here. The mesh for this simulation was generated by first meshing one surface of the thin-film layer with a reasonable initial triangular mesh and then allowing the mesher to mesh the remaining volume with tetrahedrons. This leads to a relatively small number of tetrahedrons being used to mesh the thin layer (specifically $2,300$), and consequently leads to them having very low mesh quality (the average scaled Jacobian metric illustrated in Fig. \ref{fig:mesh-quality} is $0.0274$ with a standard deviation of $0.0199$ within this thin layer). The rest of the geometry, described in detail in Fig. \ref{fig:xmon-thin}, leads to a mesh of the entire geometry containing $45,099$ tetrahedrons. 

Despite the thin tetrahedrons comprising only a minority of the entire mesh, they dominate the behavior of the solution. This is shown in Fig. \ref{fig:xmon-thin} where the input impedance of the transmon is displayed. It is seen that around $2.6 \, \mathrm{GHz}$ the field-based solver has begun to break down due to the multiscale modeling challenges and the solution data is no longer reliable at lower frequencies. These results were generated with an incomplete LU (ILU) preconditioner and GMRES iterative solver using typical settings that generally work for geometries with a similar number of mesh elements but no highly multiscale features. In principle, a more aggressive preconditioner could be formed to help offset these issues to even lower frequencies, but this is not always a practical solution, particularly for larger-scale simulations where parallelizable preconditioners need to be used. An alternative strategy is to use a more robust FEM formulation, such as TCS or the potential-based method of \cite{elkin2025improvements}, which are shown to exhibit no convergence issues in Fig. \ref{fig:xmon-thin} due to the $175 \, \mathrm{nm}$ thick layer being included in the simulation.

\begin{figure}[t!]
    \centering
    \includegraphics[width=0.9\linewidth]{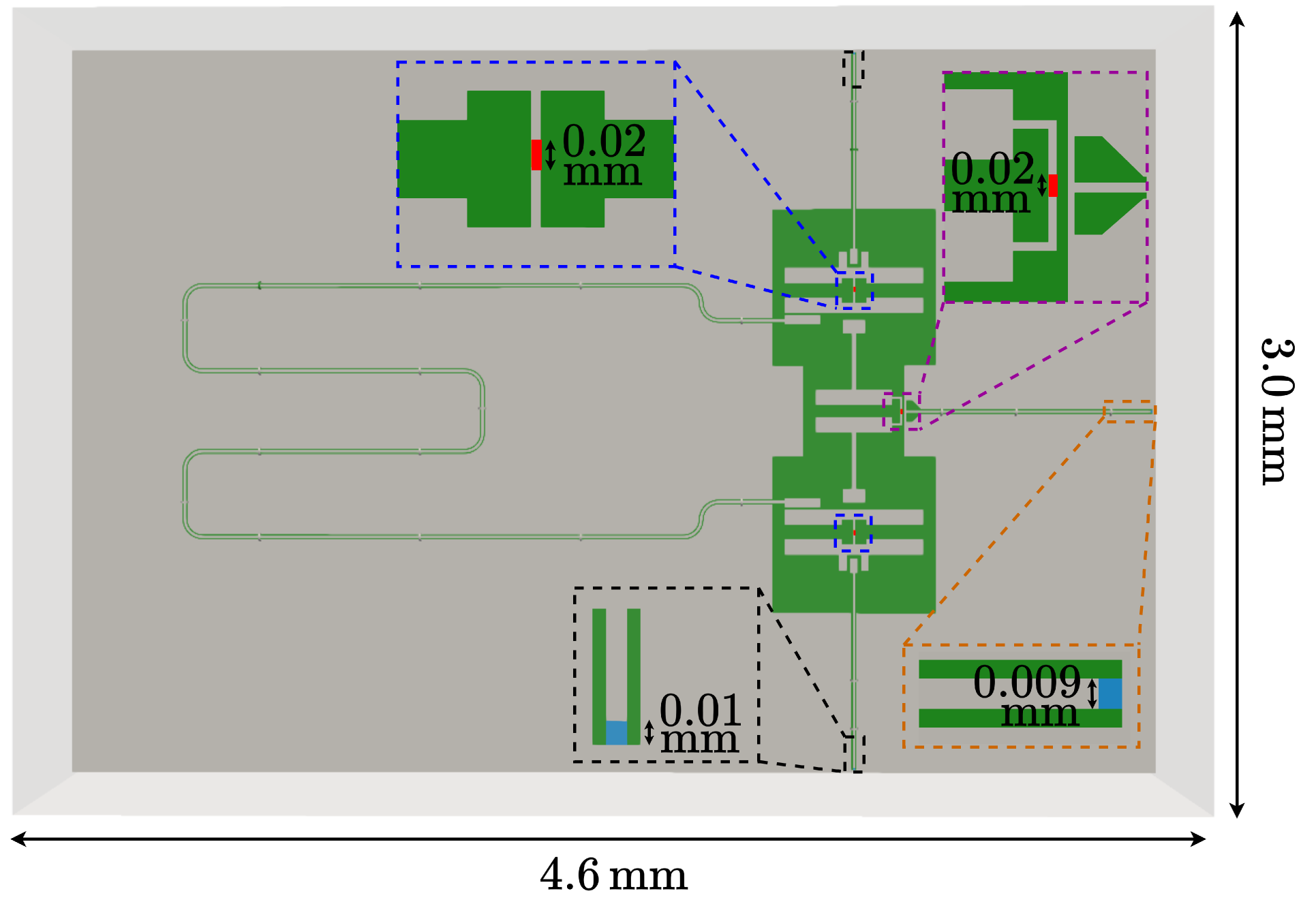}
    \caption{Two-qubit device based on \cite{mundada2019suppression} that uses a combination of a fixed resonator and tunable coupler to suppress qubit-qubit crosstalk. The substrate (green) has a relative permittivity of $\epsilon_r = 11.7$. The qubits and tunable coupler are excited through lumped ports (red) to extract the impedance parameters needed in evaluating couplings following the approach of \cite{khan2024field}. The feeding coplanar waveguide lines are also terminated with lumped ports (blue). The metal is treated as PEC to focus convergence effects purely on the geometric aspects of the device. }
    \label{fig:two-qubit-device}
\end{figure}

While this thin layer is challenging to deal with, the overall geometry is still quite simple, and so we see that TCS and the potential-based methods perform similarly. As a more comprehensive test, we now perform time domain simulations for more practical geometries. In particular, we consider the coplanar waveguide resonator shown in Fig. \ref{fig:cpw-plus-freq-response} and a more complex two-qubit device shown in Fig. \ref{fig:two-qubit-device}. This two-qubit device is based on the geometry studied experimentally in \cite{mundada2019suppression}, which couples the two qubits together through a half-wavelength coplanar waveguide resonator and a tunable coupler to suppress qubit-qubit crosstalk. Various details about the meshes of these two geometries are summarized in Table \ref{table:iter-meshes}. For each device, we perform a simulation with a modulated Gaussian input pulse that has a center frequency of $5 \, \mathrm{GHz}$ and a bandwidth of $1.325 \, \mathrm{GHz}$.

\begin{table}[t]
\caption{Statistics for the meshes used in Fig. \ref{fig:td_iter_subplots}} 
\renewcommand{\arraystretch}{1.5}
\centering
\newcolumntype{D}{>{\centering\arraybackslash} m{0.1\textwidth}}
\newcolumntype{M}{>{\centering\arraybackslash} m{0.18\textwidth}}
\footnotesize
\label{table:iter-meshes}
\begin{tabular}{M|D|D}
 \hline
 \hline
  &  \textbf{CPW \hspace{1cm} (Fig. \ref{fig:cpw-plus-freq-response})} & \textbf{Qubit \hspace{1cm} (Fig. \ref{fig:two-qubit-device})} \\
 \hline
 \hline
 Number of tetrahedrons &  $3.90 \times 10^5$ & $4.84 \times 10^5$ \\
 \hline
 Number of edges  &  $4.27 \times 10^5$ & $5.22 \times 10^5$ \\
 \hline
 Number of nodes  &  $5.50 \times 10^4$ & $6.52 \times 10^4$ \\
 \hline
 Avg. edge length (mm) &  $5.95 \times 10^{-2}$ & $3.59 \times 10^{-2}$ \\
 \hline
 Max. edge length (mm) &  $9.19 \times 10^{-1}$ & $3.25 \times 10^{-1}$ \\
 \hline
 Min. edge length (mm) &   $9.28 \times 10^{-4}$ & $1.20 \times 10^{-5}$ \\
 \hline
 Edge std. dev. (mm) &   $1.12 \times 10^{-1}$ & $4.08 \times 10^{-2}$\\
 \hline
 \hline
\end{tabular}
\end{table}

\begin{figure}[t!]
    \centering
    \includegraphics[width=0.9\linewidth]{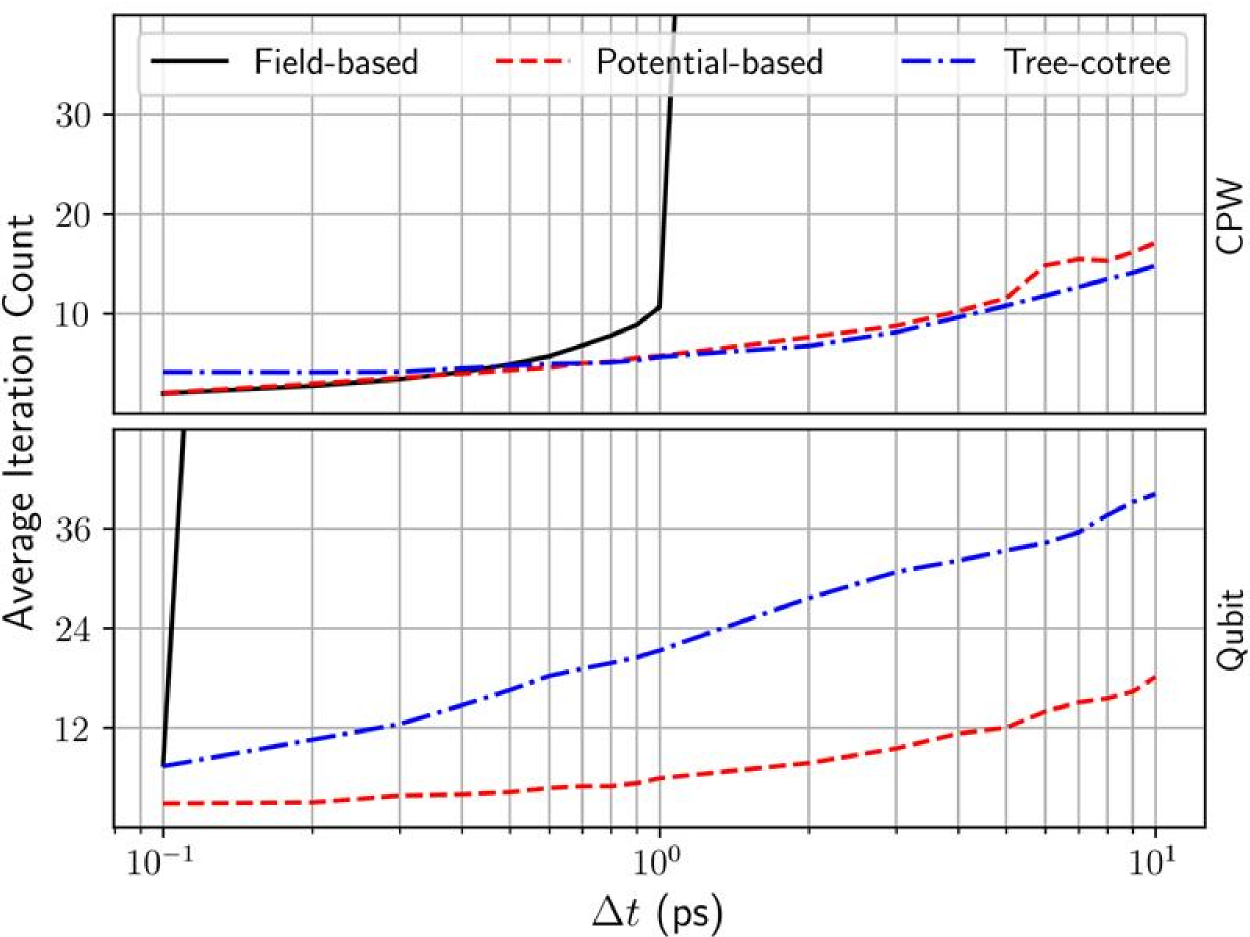}
    \caption{Average number of iterations required for GMRES to converge at each time step for the different geometries. ``CPW'' corresponds to the coplanar waveguide resonator of Fig. \ref{fig:cpw-plus-freq-response} and ``Qubit'' corresponds to the two-qubit device shown in Fig. \ref{fig:two-qubit-device}. The more multiscale features of the two-qubit device make the TCS method begin to perform worse than the potential-based method that is designed for multiscale devices.}
    \label{fig:td_iter_subplots}
\end{figure}

To compare the performance of the methods, we run $1 \, \mathrm{ns}$ duration simulations with different time step sizes $\Delta t$ and track the average number of iterations for GMRES to converge at each time step, which is shown in Fig. \ref{fig:td_iter_subplots}. The maximum time step considered is $20$ steps per period, which corresponds to the maximum time step generally considered acceptable for temporal sampling accuracy requirements. The smallest time step, corresponding to $2000$ steps per period, is selected to represent situations where enhanced accuracy may be needed or for scenarios where nonlinear or multiphysics coupling may be present. All of these time steps considered are typically orders of magnitude larger than what the stability constraint would be for these meshes, so we utilize an unconditionally stable time marching method (see Section \ref{subsubsec:fem-td}) for each solver. The time marching is performed with GMRES with 30 iterations per restart and a relative tolerance of $10^{-6}$. A standard ILU preconditioner is also used for each method.

For the coplanar waveguide resonator, the field-based formulation breaks down for $\Delta t \geq 2.0 \, \mathrm{ns}$ while TCS and the potential-based method perform comparably. However, for the two-qubit device, we see that the high number of multiscale features causes TCS to begin to perform worse than the potential-based method that is designed to be robust for such situations. In comparison to these more robust solvers, the field-based formulation breaks down for all $\Delta t \geq 0.2 \, \mathrm{ps}$. As with the previous results, more aggressive preconditioning can potentially alleviate some of these breakdown issues; however, they will still occur at some point unless a method is utilized that eliminates this non-ideal aspect of the discretized system.

Before moving on, we note that modeling multiscale geometries is also a challenge for integral equation methods. In this case, one has a similar low-frequency breakdown effect that must be grappled with, as well as what is known as the dense-mesh breakdown effect. In the case of dense-mesh breakdown, the underlying properties of the integral operators being discretized leads to a progressively ill-conditioned matrix equation as the mesh is refined more and more \cite{andriulli2008multiplicative}, making the problem increasingly challenging to be solved with iterative solvers. Addressing these fundamental challenges remains an area of active research interest, although various approaches have reached a level of maturity that they can be implemented and tested for a particular application \cite{andriulli2008multiplicative,qian2009fast,liu2018potential,zhao2000integral,taskinen2006current,vico2016decoupledPot}. One point of interest with respect to certain PEEC methods is that these do not suffer from these breakdown effects in the same way as standard MoM approaches \cite{ruehli2017circuit}, making them more resilient to these adverse effects.

It is also important to note that physics-based fast algorithms like the MLFMA have their \textit{own} breakdown effects on top of those of the basic integral operators, which must be tackled in any implementation. Again, various approaches exist for handling this for the ``standard'' integral equation formulations (e.g., see \cite{xia2018low} and the references therein), but for more bespoke approaches that are optimized for a particular application one may have to address all of these breakdown effects again in the development process, extending the overall development timeline.

\section{Outlook and Future Directions}
\label{sec:conclusion}
Throughout this work, we have reviewed the key concepts underlying the operation of the most popular CEM methods and elucidated why different aspects of modeling cQED devices can stress the limitations of many of these tools. Although research related to these kinds of challenges has been conducted for many years, the highly multiscale aspects of cQED devices can still remain a challenge to deal with, especially as the size of a device grows. As such, more research into the underlying CEM methods is still required to improve the ability of designers to efficiently model their devices. Fortunately, cQED devices are not the only emerging systems stressing these kinds of limitations in CEM tools; e.g., the trend toward heterogeneous integration of chiplets to enable advanced semiconductor devices is also leading to highly multiscale systems of large sizes that designers need to efficiently model \cite{lau2022recent}. Hence, the field of cQED modeling will likely be able to benefit from advances in CEM tools driven by other application areas in addition to their own. However, there are also many important research directions to explore specifically in the area of cQED modeling, which we discuss throughout the rest of this section.

\subsection{Unconventional Material Modeling and Computational Benchmarking}
\label{subsec:materials-and-benchmarks}
While cQED researchers may benefit from broader CEM research driven by other application areas, cQED modelers can also learn important lessons from these other applications areas to improve cQED modeling procedures. For example, one of the key activities that has been substantially invested in within the engineering communities are computational benchmarking efforts for specific problem sets of interest to different industries \cite{yilmaz2025special,dault2025computational,barnes2025benchmarking}. While the exact scope of a ``computational benchmark'' can take different forms, the main components will ideally involve a set of well-defined, publicly-releasable test problems with realistic geometries and corresponding measurement and simulation results. Of key importance to the success of a benchmarking problem is the collaboration between the teams handling the fabrication, measurement, and simulations to carefully minimize as many uncertainties in the problem definition as possible. This typically requires many additional independent measurement and characterization steps beyond what would be done to test a device for typical experiments to aid in determining the differences between designed values and the actual fabricated properties (e.g., dimensions, material properties, surface roughness, etc.). Although the work involved can be significant, such activities have been key to industrial efforts being able to achieve excellent agreement between measurements and simulations for applications like advanced semiconductor devices \cite{barnes2025benchmarking}. These benchmarks can also be designed for different levels of simulation difficulty, with simple problems helping new researchers entering the field learn best practices and more difficult ``challenge problems'' helping test state-of-the-art tools as well as identify opportunities for further improvements. 

Although some grassroots efforts have begun in this vein \cite{shanto2024squadds}, there are significant opportunities for a more systematic and multi-institutional process to be initiated in this area. One area of focus would be the design of benchmark devices to better disentangle the effects of dielectric permittivity and kinetic inductance. Many researchers attempt to determine these parameters from their devices by simply modifying the properties in their simulations until a resonant frequency aligns with measurements. However, these material properties can compensate for one another, and the broader ``fitting'' process being followed can easily hide systematic simulation inaccuracies by ``forcing'' the model to agree with the measurements. Further, given the simple manner in which permittivity and kinetic inductance influence the phase velocity, primarily looking at resonator frequencies to ``determine'' these parameters could also lead to incorrect conclusions on whether a modeling procedure is an accurate way to incorporate unconventional materials like thin superconductors into CEM models.

For instance, some resources suggest considering an infinitely-thin surface with a prescribed surface inductance \cite{kerr1996surface,kongpop2018modeling,whitaker1988propagation}, while others suggest using a volumetric discretization with an imaginary-valued conductivity \cite{Levenson-Falk_2025}. In each case, the values of the surface inductance or imaginary-valued conductivity are often computed from idealized solutions to the London equations while usually only considering the London penetration depth as an input parameter. Invariably, these procedures often resort to simply choosing the London penetration depth as a fitting parameter until the simulation results match a measured resonant frequency. More involved approaches have also been explored that utilize Mattis-Bardeen theory and additional input parameters to compute the EM properties of the superconductor \cite{lopez2025superconducting,duan2025simulation}, but these still are generally operated from the perspective of attempting to match measured resonant frequencies rather than looking at a wider range of QoIs. Given that these procedures can lead to a penetration depth of similar magnitude to the thickness of the superconductors \cite{lopez2025superconducting}, conventional wisdom would suggest that volumetric discretization of the superconductor should be important to achieve accurate simulation results for a wide range of QoIs (e.g., for resonant frequencies, resonator quality factors, flux linkage to superconducting loops, etc.). However, given the severe ill-conditioning that would come with a volumetric discretization of a $O(100 \, \mathrm{nm})$ thick superconductor using conventional simulation methods, these approaches would benefit from being tested with a more careful computational benchmarking effort and CEM methods that are robust for multiscale structures.

Another area that could benefit from systematic computational benchmarking efforts is modeling the effects of loss in cQED devices. Many simulation studies have significant difficulties achieving good agreement between simulated and measured resonator linewidths across a device using a single set of material properties, and it remains an open question exactly why this is the case. More generally, modeling high-Q resonances accurately is a consistent challenge in CEM tools, with specialized methods sometimes required to improve the performance of the underlying model, as well as in other auxiliary procedures like interpolation methods for fast frequency sweeps or in adaptive mesh refinement algorithms \cite{yuan2022method,bao2025automatic}. Correspondingly, it is difficult to assess without a careful benchmarking effort if the issue lies in the CEM method itself or in the model inputs agreeing with the actual fabricated device properties.

Further opportunities also exist in characterizing the qubits themselves. Most cQED simulation models consider the qubits in terms of only a few parameters: e.g., the Josephson energy and the charging energy for a transmon \cite{koch2007charge,roth2022transmon}. Some recent efforts have identified a likely shortcoming of this approach, especially in achieving good agreement for transition frequencies that involve more than the first three energy levels \cite{willsch2024observation}. In the area of integrated circuit modeling (e.g., for RF amplifiers), models with $O(100)$ parameters for each transistor are usually required, suggesting that a more detailed set of parameters will also likely be needed to design suitable compact models of cQED qubits. Of course, research to effectively characterize a fabrication process so that designers can be provided with this broader set of robust input parameters would also be required.

\subsection{Advances to CEM Methods}
\label{subsec:cem-advances}
There are also significant opportunities for research purely on the CEM methods themselves. As discussed earlier, there is still a need to improve the underlying CEM methods by making them more robust for the highly multiscale geometries commonly encountered in cQED devices. Likewise, strategies like domain decomposition methods also need to be reconsidered in the context of highly multiscale geometries. The presence of these multiscale features can worsen the conditioning of the global interface problem that needs to be solved for frequency domain methods, or can cause challenges related to stability constraints if a time domain method is used \cite{jin2015finite}. Hence, new domain decomposition formulations or preconditioning strategies should be investigated. If large numbers of eigenpairs will need to be extracted to model larger cQED devices, more research on domain decomposition methods specialized for eigensolvers and the use of time domain methods for extracting eigenpairs will be beneficial.

Another important but challenging task is the accurate and efficient modeling of surface participation ratios. The use of singular basis functions in FEM or MoM codes could greatly aid in this, but for these approaches to be effective, the user usually needs to have some prior knowledge of the kind of singularity expected to be present \cite{peterson2016basis}. Hence, a better understanding of the asymptotic scaling of singularities for thin superconducting geometries would be useful in these studies. Alternatively, cQED modeling could benefit from exploring the use of more advanced adaptive mesh refinement algorithms, such as adjoint-based goal-oriented mesh refinement that can focus the refinement effort specifically towards difficult to simulate QoIs \cite{becker2001optimal,harmon2020adjoint,wang2023posterior}. Beyond modeling qubit loss, finding strategies to efficiently account for these similar kinds of loss mechanisms over the length scale of transmission line resonators would also be valuable.

For most cQED simulations, FEM is the default tool that has been used. Integral equations represent a powerful alternative that, with sufficient development, can provide substantial efficiency boosts for the kinds of structures commonly used with cQED devices. For instance, specialized integral equation implementations for layered media with coplanar waveguide layers could dramatically reduce the number of basis functions needed to simulate large devices \cite{michalski2002multilayered,okhmatovski2024theory,drissi1991analysis,dib1991theoretical,wu1995full}. Integral equation approaches have also been very successful in creating flexible reduced order models in which many smaller basis functions are combined into larger ``macro basis functions'', usually for geometric features that are repeated many times in an overall geometry, but this is not strictly required \cite{prakash2003characteristic,garcia2008iterative,gonzalez2011interpolatory}. Applications of these approaches to date have primarily been for simulating radiation or scattering from antennas or other periodic structures (e.g., metasurfaces), but these strategies could potentially be adapted to greatly reduce the number of basis functions needed to model qubits within large devices. Other reduced order models for sections of resonators that are repeated often in a large device could also be sought. 

\subsection{Advances to cQED Theory and Device Design}
\label{subsec:cQED-advances}
In addition to improving the CEM methods themselves, significant research is also needed on cQED theory and design procedures to determine alternative strategies for using CEM tools more efficiently. For example, while cQED theory is very convenient to express theoretically in terms of eigenpairs, actually computing these eigenpairs for 3D geometries is substantially difficult. Further, because traditional engineering applications do not often employ CEM eigensolvers, these tools are much less likely to be significantly advanced by the broader research community or in commercial tools. 

Instead, CEM advances that will happen from other communities will most likely be focused on driven simulations that can be used to compute impedance, scattering, or other network parameters. Some research has focused on extracting important Hamiltonian parameters directly from impedance (or other network) parameters \cite{houck2008controlling,solgun2019simple,roth2022full,solgun2022direct,khan2024field,labarca2024toolbox}, but more research in this vein is needed to continue expanding what aspects of a device can be characterized in this more scalable way. Likewise, the use of multiphysics modeling methods for qubit control and readout should also continue to be explored \cite{roth2024maxwell,2024_Elkin_JTWPAs,elkin2025ims,elkin2025multiphysics}. Connecting the results of other multiphysics simulations (e.g., EM and thermal) to understand heat dissipation and cooling throughout a device and how this connects to noise generation that can be fed into open quantum system models could also be of interest \cite{simbierowicz2024inherent}. 

Another emerging trend at the intersections of CEM and the design of cQED based quantum computer is to maximize the translational symmetry of the qubit circuitry on the coherent chip.  First, a unit cell containing a single qubit or a few qubits along with their control, readout, and coupling structures, is designed to achieve the desired ``local'' performance metrics (qubit frequencies, coupling capacitance and inductance values, etc.). This unit cell is then repeated along two perpendicular directions in the plane, thus generating a large, periodic array \cite{tamate2022toward}. In order to preserve the periodicity and to accommodate complex routing, the vertical dimension is increasingly being used for routing signal and control lines to and from the coherent chip using, e.g., pogo pins, micro-bumps, or through-silicon vias (TSVs) \cite{hazard2023characterization, spring2025fast}. The great advantage of this approach is its highly scalable nature, as it circumvents the difficulty of ``fanning in'' a large number of wires in-plane from the chip edge to an ever-increasing number of qubits in the middle.  (In flip-chip architectures, this difficulty is relegated to a carrier chip but is no less severe.)  More germane to the present context, such a design also offers the possibility of significantly reducing the burden on the CEM modeling. Specifically, one can now generate and refine a mesh for a single unit cell, perform an initial verification of qubit design targets by running a cheap unit-cell simulation, and then use periodic boundary conditions (PBCs -- also known as Floquet boundary conditions in some CEM tools \cite{jin2015finite,munk2005frequency}) in the plane to simulate the entire chip much more efficiently. With special handling for the necessarily finite extent of the 2D structure, this ``design for simulatability'' approach promises to provide comparable accuracy with a significantly reduced computational cost \cite{mohseni2024build}.  This can appreciably accelerate the design iteration cycle, as well as enabling the modeling of chips with hundreds or thousands of qubits with much more limited computational resources.

Obviously, many more opportunities also exist for this burgeoning field. However, from a modeling perspective, more focus should be placed on the scalability of the overall computational approach being proposed. This will be critical to enable computational methods to better support the scaling of cQED devices.

\subsection{Conclusion}
\label{subsec:conclusion}
Overall, while CEM is a mature field, there are still significant opportunities to engage in further research on these methods to meet the challenges of emerging applications such as modeling cQED devices. More collaborations between cQED and CEM researchers will help propel these efforts forward and accelerate when improved modeling tools can begin to be used to aid in the predictive design of increasingly complicated devices. Overcoming these challenges will be an important piece in the overall ecosystem to improve the performance of cQED technologies, helping these technologies achieve utility for a wide range of exciting applications in the years to come.

\begin{acknowledgments}
S. T. Elkin, G. Khan, and T. E. Roth gratefully acknowledge support from the National Science Foundation under Grant No. 2202389 and by a gift from Google LLC. This work was produced by Fermi Forward Discovery Group, LLC under Contract No. 89243024CSC000002 with the U.S. Department of Energy, Office of Science, Office of High Energy Physics. Publisher acknowledges the U.S. Government license to provide public access under the DOE Public Access Plan.
\end{acknowledgments}


\bibliography{main}

\end{document}